\documentclass[aps,superscriptaddress,nofootinbib,showpacs,eqsecnum,preprint,tightenlines]{revtex4}
\usepackage{hyperref}
\usepackage{epsfig,rotating}
\usepackage{amsmath,amssymb}
\usepackage{dsfont}
\usepackage{bbm}
\usepackage{slashed}
\usepackage{slashbox}
\numberwithin{equation}{section}

\newcommand{\be}{\begin{equation}}
\newcommand{\ee}{\end{equation}}
\newcommand{\bea}{\begin{eqnarray}}
\newcommand{\eea}{\end{eqnarray}}

\newcommand{\vx}{\vec{x}}

\newcommand{\vp}{\vec{p}}

\newcommand{\vq}{\vec{q}}
\newcommand{\vQ}{\vec{Q}}
\newcommand{\vk}{\vec{k}}

\begin{document}

\title{Cosmological Implications of Light Sterile Neutrinos produced after the QCD Phase Transition}

\author{Louis Lello}
\email{lal81@pitt.edu}

\author{Daniel Boyanovsky}
\email{boyan@pitt.edu}

\affiliation{Department of Physics and Astronomy, University of Pittsburgh, Pittsburgh, PA 15260, USA}

\date{\today}

\begin{abstract}

  We study the production of sterile neutrinos in the early universe from $\pi \rightarrow l \nu_s$ shortly after the QCD phase transition in the absence of a lepton asymmetry while including finite temperature corrections to the $\pi$ mass and decay constant $f_{\pi}$. Sterile neutrinos with masses $\lesssim 1 MeV$ produced via this mechanism freeze-out at $T_f \simeq 10 MeV$ with a distribution function that is highly non-thermal and features a sharp enhancement at low momentum thereby making this species \emph{cold} even for very light masses. Dark matter abundance constraints from the CMB and phase space density constraints from the most dark matter dominated dwarf spheroidal galaxies provide upper and lower bounds respectively on combinations of mass and mixing angles. For $\pi \rightarrow \mu \nu_s$, the bounds lead to a narrow region of compatibility with the latest results from the $3.55 \mathrm{KeV}$ line. The non-thermal distribution function leads to free-streaming lengths (today) in the range of $\sim \mbox{few kpc}$ consistent with the observation of cores in dwarf galaxies. For sterile neutrinos with mass $\lesssim 1 eV$ that are produced by this reaction, the most recent accelerator and astrophysical bounds on $U_{ls}$ combined with the non-thermal distribution function suggests a substantial contribution from these sterile neutrinos to $N_{eff}$.
\end{abstract}

\pacs{95.35.+d,95.30.Cq,13.15.+g}

\maketitle

\section{Introduction}

The current paradigm in cosmology is that the energy content of the universe is divided into the particle species of the standard model, an unknown dark energy driving the current expansion of the universe and an unknown (cold) dark matter species ($\Lambda CDM$) \cite{book1}. Dark matter (DM) is thought to be in the form of cold thermal relics with interaction cross sections on the order of weak interaction strength (WIMPs) \cite{wimpreview} with alternate theories favoring axions \cite{axionreview} or new neutrino species \cite{sterilereview}. The standard cold dark matter cosmology explains much of the observational data yet some problems at small scales remain unexplained.

Cold dark matter N body simulations predict that dark matter dominated galaxy profiles feature a cusp, but observations suggest that the profiles are cores (core v cusp problem) \cite{corecusp,dalcantonhogan}. Additionally, simulations of $\Lambda CDM$ show that dark matter subhaloes in the Milky way are too dense for the observed satellites (too big to fail) \cite{toobig}. Both of these problems could be alleviated if the dark matter candidate is allowed to be "warm" (WDM) \cite{wdm1,wdm2,wdm3,wdm4,wdm5,wdm6}, one such candidate being a massive "sterile" neutrino \cite{warmdm,dodwid,dodwid2,abazajian3,sterilesexperiment}. The free streaming length, $\lambda_{fs} = 2\pi/k_{fs}$, is the scale that cuts off the power spectrum of density perturbations. CDM features very small ($\lesssim$ pc) $\lambda_{fs}$ which leads to cuspy profiles while WDM features $\lambda_{fs} \sim \mbox{few kpc}$ possibly explaining the observed cores. $\lambda_{fs}$ is determined by the distribution function at freeze out. Alternatively, decaying DM candidates, such as WIMPs or gravitinos, could also lead be a simultaneous solution to both of these problems \cite{kaplinghat}.

Additionally, with the discovery of neutrino masses, a considerable experimental effort has shed light on the parameters of the neutrino sector  \cite{pdg,neutrinoexperiments}. The last of the mixing angles describing neutrino oscillation has been measured and there are proposals for new facilities to probe CP violation, Dirac/Majorana nature, inverted/normal hierarchy in the active neutrino sector \cite{upcomingexperiments}. There are also some persistent short baseline anomalies (LSND, MiniBooNE) \cite{lsnd,miniboone} that can be explained with an additional sterile neutrino species \cite{sterilesexperiment} but tension exists with other experiments \cite{nosterilesexperiment}. There are plans to search for these sterile neutrinos in forthcoming experiments, many of which involve neutrino production from the decay of meson parent particles, processes in which the subtleties of the decay event itself may prove useful \cite{decay}. Other proposed experiments could search for sterile neutrinos  via modifications to oscillation formulae on short baseline experiments \cite{lello1} , monochromatic peaks searches \cite{lello1,shrock} or as contributions to lepton flavor violation  experiments  \cite{lello2}. A review of the motivation for sterile neutrinos from terrestrial experiments and a summary of some of the proposed experiments that will look for sterile neutrinos can be found in \cite{lasserre}. The latest limits on sterile neutrino mixing from atmospheric neutrino data have been set by the Super Kamiokande experiment \cite{superk} which sets the limits $|U_{\mu 4}|^2 < 0.041$. Similar bounds have been by the Daya Bay collaboration \cite{dayabay} and the analysis in \cite{giunti2,mirizzi} examines the global fits for various light sterile neutrino scenarios (3+1,3+2,3+1+1). A summary of the light sterile neutrino bounds for active-sterile mixing from accelerators, cosmology and other experiments are summarized concisely in figs 1-3 of ref \cite{kusenko2} while those for heavy steriles can be found in \cite{kusenko3}.

Several extensions of the standard model include sterile neutrino species, for instance \cite{shaposh} describes a model which is an extension of the $\nu MSM$ and purports to describe inflation, dark matter, the baryon asymmetry and neutrino oscillations. For most treatments of sterile neutrino dark matter, a nonthermal distribution function is needed in order to evade cosmological bounds \cite{planck}. Ref \cite{giunti} argues that short baseline inspired steriles (1eV) could not be in thermal equilibrium in the early universe but can be made compatible with observations by allowing the sterile to decay into very light particles. The mechanism of sterile neutrino production in the early universe through oscillations was originally studied in a body of work by Barbieri, Dolgov, Enqvist, Kainulainen and Maalampi (BDEKM) \cite{dolgovenqvist} and, in \cite{dodwid}, sterile neutrinos are argued to be a viable warm dark matter candidate produced out of LTE via the BDEKM mechanism (Dodelson-Widrow, DW). In \cite{shifuller}, light keV sterile neutrinos are produced by resonant MSW conversion of active neutrinos, similiar to DW but with resonant oscillation in the presence of a lepton number asymmetry (Shi-Fuller, SF). Models in which a standard model Higgs scalar decays into pairs of sterile neutrinos at electroweak energy scales (or higher) have also received attention \cite{boyan1,kusenko1,petraki}. Ref \cite{petraki} calculates the free streaming length and phase space density of sterile neutrinos from Higgs-like decays, both in and out of equilibrium, which is used to compare to small scale structure observations. These types of mechanisms have inspired work on understanding properties of more general nonthermal dark matter such as \cite{boyan4,boyan3}.

Recently, a signal of 3.5 keV line has been claimed at $3\sigma$ detection from the XMM Newton x-ray telescope which could be a hint of a 7 keV sterile neutrino \cite{bulbul,boyarsky}. The interpretation of the anomalous line as a signal of a sterile neutrino has been challenged \cite{jeltema,malysh} motivating further studies of the signal. In refs \cite{abazajian0,abazajian,abazajian2}, the parameter space for SF type steriles that could be compatible with the 3.5 kev signal is explored. Besides the 3.5 keV line, other observational clues seem to favor or disfavor the various mechanisms. Ref \cite{merle} claims that high redshift quasar Ly $\alpha$ signals disfavor both DW and SF mechanisms but is consistent with scalar decay.
Radiative decays of sterile neutrino dark matter candidates is constrained by the Chandra X-ray spectrum which places limits on sterile mass (for DW) at $m< 2.2keV$ \cite{watson}. Observations of dwarf spheroidal phase space densities and X-ray data in the local group essentially rule out DW steriles but still allow for SF or other mechanisms \cite{kaplinghat2}. The effects of massive neutrinos on the Sachs Wolf plateau and CMB fluctuations have been calculated and limits placed on the mass and lifetime \cite{kaplinghat3} while phase space densities of dwarf spheroidals lead to bounds a WDM sterile candidate at $m\lesssim \mbox{ few  keV}$ \cite{destri}.

The prospect of keV WDM sterile neutrinos remains an active area of investigation experimentally and theoretically. Ref \cite{kusenko1} claims keV neutrino DM produced via Higgs decays matches the bounds of small scale structure and X-ray observations while simultaneously explaining pulsar kicks. It has also been suggested that SF type steriles reproduce the appropriate galaxy distribution and could potentially lead to a test of the quark-hadron transition \cite{abazajian2}. Ranges of masses and mixing for both DW and SF mechanisms include constraints from supernovae, BBN and decay limits which can be found in \cite{abazajian3}. One of the observational windows towards the detection of light ($m \lesssim \mbox{eV}$) sterile neutrinos are from cosmological measurements of $N_{eff}$, the sum of neutrino masses and the lepton asymmetry and BBN \cite{kirilova} \cite{steigman}. A comparison of how various dark radiation sources contribute to these measurements can be found in \cite{archid}.

Ref \cite{fullkuse} considers heavy sterile neutrinos (100-500 MeV) in thermal equilibrium but decay nonthermally and finds a range of parameter space in which these models can contribute to $N_{eff}$ without violating the bounds. A mechanism of neutrino reheating in ref \cite{boehm} considers other particles which remain in local thermodynamic equilibrium (LTE) with neutrinos and decouple before photon decoupling, changing the neutrino to photon temperature ratio. Contributions to $N_{eff}$ from decaying non-thermal particles can mimic sterile neutrinos where higher moments of the distribution functions would be required to discriminate between scenarios \cite{hasen}. Its been shown that delaying neutrino freeze out contributes to dark radiation \cite{birrell} and, additionally, freeze out of Bose or Fermi degrees of freedom during QCD phase transition would lead to changes in dark radiation measurements \cite{birrell2}. Additionally, the relic densities of sterile neutrinos depend on the QCD transition and, if detection and study of these particles were possible, could offer a window to the QCD phase transition \cite{abazajian}.

To the best of our knowledge, the mechanism which is used to produce neutrinos in many terrestrial experiments, $\pi \rightarrow \mu \nu$, has not been addressed in a cosmological setting. The difficulty in such a problem is reflected in the challenges inherent to the QCD era of the early universe. The QCD phase transition, when the universe cools enough for free quarks and gluons to hadronize, continues to be an epoch in cosmology which remains to be fully understood \cite{qcdptreview}. Recently, the latest lattice QCD calculations have suggested that the QCD phase transition is continuous with a crossover at $T=155 MeV$ \cite{hotqcd}. It is generally accepted that $\pi$ mesons, the lowest lying QCD bound states, will be produced in abundance and this has motivated thorough studies of pions near the QCD phase transition. Near the phase transition, stable long wavelength pion excitations are developed which may be detectable signatures in heavy ion colliders \cite{boyan5,rajawilc}. At temperatures below the QCD phase transition, finite temperature corrections to the pion mass and decay constant become important and non-trivial \cite{nicola1}. These corrections have been studied in linear \cite{tytgat1} and non-linear sigma models\cite{jeon}, using QCD sum rules at finite temperature \cite{dominfete}, hidden local symmetry models \cite{harada} and chiral perturbation theory \cite{nicola2}. \\

\textbf{Goals:} The main goal of this work is to understand the production and freeze out of sterile neutrinos from $\pi \rightarrow l \nu_s$ shortly after the QCD transition. With the finite temperature corrections to the pion mass and decay constant, it is possible to consider the quantum kinetics of sterile neutrinos that are produced in the early universe from the same mechanisms which are employed by land based accelerator experiments, namely $\pi \rightarrow l \nu_s$. We obtain the distribution function of a sterile neutrino produced from pion decay in the early universe by including finite temperature corrections and investigate the immediate observational consequences. We will be restricting our attention to the study of \emph{light} sterile neutrinos with masses $m_{\nu} \lesssim 1 MeV$. These will be shown to freeze out while they are still relativistic with non-thermal distributions.

\begin{itemize}
\item With a non-thermal distribution function, measurements of $\Omega_{DM}$ give an upper bound for the energy density of the sterile neutrinos today. A complementary bound is obtained by considering the velocity dispersion and energy density of dwarf spheroidal galaxies. These measurements coupled with the non thermal distribution place bounds on combinations of masses and mixing matrix.

\item The free streaming length, which is small for cold dark matter candidates and larger for warmer dark matter candidates, is dependent on the specific form of the distribution function. We obtain $\lambda_{fs}$ from the non-thermal distribution function arising from pion decay.

\item A light sterile neutrino of $m \lesssim \mbox{1 eV}$ could be relativistic at the time of matter-radiation equality and potentially contribute to the measurement of $N_{eff}$. We investigate the contribution to this number from the pion-produced sterile neutrino and how the equation of state parameter, w, evolves from relativistic to non-relativistic compared to a thermal distribution.

\end{itemize}

\textbf{Brief Summary of Results:}

\begin{itemize}

 \item We find the \emph{non-thermal} distribution function for sterile neutrinos that were produced via pion decays shortly after the QCD phase transition. This distribution features a low momentum enhancement similar to that found in resonantly produced models (Shi-Fuller). A key difference between the two models is that resonant model requires a non-zero lepton asymmetry which is absent in the distribution that we obtain. This mechanism produces a colder sterile neutrino dark matter candidate, similar to MSW resonance enhancement, but without the requirement of a lepton asymmetry. A calculation of the equation of state shows that, while freeze-out occurs as the particles are still relativistic, this type of sterile neutrino becomes non-relativistic very quickly, namely when $T \sim m$, as opposed to thermal distributions which become non-relativistic when $T \ll m$.

\item We obtain bounds on combinations of sterile neutrino mass and mixing matrix elements from CMB observations and dark matter dominated galaxies. Using the observed dark matter density from Planck as an upper bound for the sterile neutrino energy density leads to an upper bound on a combination of the mass and mixing matrix:

\be
m_{\nu_s} \frac{|U_{\mu s}|^2}{10^{-5}} \le 0.739 \, \mbox{keV} ~~;~~ m_{\nu_s} \frac{|U_{e s}|^2}{10^{-5}} \le 7242 \, \mbox{keV} \,.
\ee A complementary bound is obtained from the primordial phase space density and compared to present day observations of dark matter dominated galaxies. By requiring that the primordial phase space density of sterile neutrinos be larger than the observed density and velocity dispersion relations for dark matter dominated galaxies leads to a lower bound on a different combination of mass and mixing matrix:

\be
m_{\nu} \left(\frac{|U_{\mu s}|^2}{10^{-5}}\right)^{1/4} \ge 0.38 \, \mbox{keV} ~~;~~ m_{\nu} \left(\frac{|U_{e s}|^2}{10^{-5}}\right)^{1/4} \ge 6.77 \, \mbox{keV} \,.
\ee The 7.1 keV sterile neutrino predicted by \cite{bulbul,boyarsky} (with $|U|^2 = 7*10^{-11}$) is consistent with these bounds for sterile neutrinos produced from $\pi \rightarrow \mu \nu_s$ within a narrow region but not from $\pi \rightarrow e \nu_s$.

\item To be a suitable dark matter candidate, the free streaming length must be smaller than the size of the dark matter halo. The free streaming length is calculated using the non thermal distribution function and, due to the enhancement at low momentum, is reduced for $keV$ type steriles. The free streaming length today is given by

\be
\lambda_{fs}^{\mu} (0) \sim 7.6 \, \mbox{kpc} \left(\frac{\mbox{keV}}{m_{\nu}}\right)         ~~;~~ \lambda_{fs}^e(0) \sim 16.7 \, \mbox{kpc} \left(\frac{\mbox{keV}}{m_{\nu}}\right)
\ee A sterile species that is still relativistic at the time of matter-radiation equality will contribute to $N_{eff}$ and, since this type of sterile neutrino becomes non-relativistic at $T \sim m$, the contributions to $N_{eff}$ are only valid for $m_{\nu} \lesssim 1 eV$. Parameterizing the contribution to dark radiation as $N_{eff} = N_{eff}^0 + \Delta N_{eff}$ where $N_{eff}^0 = 3.046$ is the standard model contribution \cite{neffboltz}, the sterile neutrinos we consider here contribute

\be
 \Delta N_{eff}\Big|_{\pi \rightarrow \mu \nu} = 0.0040 *\frac{|U_{\mu s}|^2}{10^{-5}} ~~;~~ \Delta N_{eff} \Big|_{\pi \rightarrow e \nu} = 9.7 *10^{-7} \frac{|U_{e s}|^2}{10^{-5}} \,.
\ee Combining with a recent analysis \cite{dayabay,superk} we find that $\Delta N_{eff} \lesssim 4$, suggesting that this mechanism could provide a significant contribution to $N_{eff}$ although severe tensions remain between accelerator/reactor fits and CMB observations.

\end{itemize}

\section{Dynamics of decoupled particles} \label{decoupleddynamics}

In this section we gather the general essential ingredients for several cosmological quantities in terms of the distribution function of the dark matter particle. Kinetic theory in a cosmological setting is well understood \cite{book2,book3,debbasch}, the purpose of this section is to review the details of the dynamics of decoupled particles which will be relevant for the following sections. The results of this section will be used in conjunction with the distribution obtained from quantum kinetics to place limits on sterile neutrino parameters.

For flat Friedmann-Robertson-Walker (FRW) cosmologies, particles follow geodesics described by

\be
ds^2 = dt^2 - a(t)^2 d\vx^2 \,.
\ee The only non-vanishing Christoffel symbols are given by

\be
\Gamma^{i}_{j 0} = \Gamma^{i}_{0 j} = \frac{\dot{a}}{a} \delta^i_j ~~;~~ \Gamma^0_{i j} = \dot{a} a \delta_{i j} \, .
\ee The geodesic equations are then given by

\be
\dot{q}^0 =-\frac{a^2 H \vq^{\,2}}{q^0} ~~;~~ \dot{\vq} = -2H \vq
\ee where $q^{\mu} = dx^{\mu}/d\lambda$ and $\lambda$ is an affine parameter. The solution is given by

\be
\vq = \frac{\vq_c}{a^2}
\ee where $\vq_c$ is a constant comoving momentum. The geodesics of massive particles imply $g_{\mu \nu} q^{\mu} q^{\nu} = m^2$, leading to the dispersion relation $q^0 = \sqrt{m^2+a^2 \vq^{\,2}}$.

The physical energy and momentum is that which is measured by an observer at rest with respect to the expanding spacetime. The stationary observer is one who measures with an orthormal tetrad

\be
g_{\mu \nu} \varepsilon^{\mu}_{\alpha} \varepsilon^{\nu}_{\beta}= \eta_{\alpha \beta} = diag(1,-1,-1,-1)
\ee or

\be
\varepsilon^{\mu}_{\alpha} = \sqrt{|g^{\mu \alpha}|} \, .
\ee With this, the physical energy/momentum are given by

\be
E = g_{\mu \nu} \varepsilon^{\mu}_0 q^{\mu} = q^0 ~~;~~ Q_f = g_{\mu \nu} \varepsilon^{\mu}_i q^{\nu} = a q^i = \frac{q^i_c}{a} \,.
\ee

The buildup of the distribution function arises from a Boltzmann equation in which decaying particles source the equation. Provided that any other interactions can be neglected, such as a sterile neutrino's interaction with standard model particles, and that the distribution is isotropic, then the kinetic equation is given by

\be
\frac{df}{dt}(Q_f,t) = \frac{\partial f}{\partial t} - H Q_f \frac{\partial f}{\partial Q_f} = \mathbbm{P}[f]
\ee where $\mathbbm{P}$ is the production integral which will be discussed in a subsequent section. Upon freeze out, the production integral vanishes and the distribution function follows geodesics governed by a collisionless Liouville equation, namely with $\mathbbm{P} = 0$. We denote the decoupled distribution as $f_d$ to distinguish it from the full distribution which is explicitly a function of time. It is easy to see that a solution for the decoupled distribution (with $\mathbbm{P} = 0 $) are functions of the form

\be
f_d(Q_f,t) = f_d (a(t) Q_f) = f_d(q_c)
\ee which depends on the scale factor through the comoving momentum.

For this type of distribution function, not necessarily thermal, the kinetic stress-energy tensor is given by

\be
T^{\mu}_{\nu} = g \int \frac{d^3 Q_f}{(2\pi)^3} \frac{q^{\mu} q_{\nu}}{q^0} f_d(q_c)
\ee where $g$ is the internal degrees of freedom of the particular species. The number density, energy density and pressure are obtained in a straightforward manner as

\bea \label{thermoquantities}
 n =  g \int \frac{d^3Q_f}{(2\pi)^3} f_d(q_c)  & ; & \rho  =  T^0_0 = g \int \frac{d^3Q_f}{(2\pi)^3} \sqrt{Q^2_f + m^2} f_d(q_c) \\
 T^i_j = -\delta^i_j \frac{g}{3} \int \frac{d^3Q_f}{(2\pi)^3} \frac{|\vq_c|^2}{E_q} f_d(q_c)  & \rightarrow & \mathcal{P} = \frac{g}{3} \int \frac{d^3Q_f}{(2\pi)^3} \frac{|\vQ_f|^2}{\sqrt{Q_f^2 +m^2}} f_d(q_c) \, .
\eea Then, introducing the photon energy density today, we can write the contribution to the energy density as

\be \label{omegadm}
\Omega h^2 = \frac{\rho h^2}{\rho_{crit}} = \frac{h^2 n_{\gamma}}{\rho_{c}} \frac{\pi^2 \rho}{2 \zeta(3) T_{\gamma}^3} \,.
\ee The average momentum squared per particle is given by

\be
\overline{ \vQ^2 } = \frac{\int \frac{d^3Q_f}{(2\pi)^3} \vQ_f^2 f_d(q_c) }{\int \frac{d^3Q_f}{(2\pi)^3} f_d(q_c)} \,.
\ee For a nonrelativistic species this is related to the average velocity per particle via $\overline{ \vQ^2}  = m^2 \overline{\vec{V}^2}$ and to the pressure/energy density as will be discussed shortly. The Hubble factor in a radiation-dominated cosmology is given by

\be \label{hubble}
H(t) = 1.66 \, \frac{g(T)^{1/2} T(t)^2}{M_p} \,. \ee

Since the distribution function after freeze-out obeys the Liouville equation, it is straightforward to verify that the number density and energy density obey a continuity equation

\be \label{numbercons}
\frac{dn}{dt} + 3 H(t) n(t) = 0 ~~;~~ \frac{d \rho}{dt} + 3 H(t)(\rho(t) + \mathcal{P}(t)) =0 \,.
\ee The entropy density for an arbitrary distribution function is given by

\be
s_d(t) = -g\int \frac{d^3 q_f}{(2 \pi)^3} \Bigg[ f_d \ln f_d \pm (1 \mp f_d) \ln (1\mp f_d) \Bigg]
\ee where the upper (lower) is for fermions (bosons). For frozen distribution functions, ie one obeying a collisionless Liouville equation, we have a another continuity equation

\be
\frac{ds}{dt} +3 H(t) s(t) = 0 \, .
\ee This gives the result that the comoving entropy density, $sa^3$, is constant.

With a mixture of several types of species in LTE and additional non-thermal species with entropy $s_d$, entropy conservation gives

\be \label{entropycons}
\left[ \frac{2 \pi^2}{45} g(T) T_{\gamma}^3 + s_d \right] a^3(t) = const \, .
\ee where $T_{\gamma}$ is the photon temperature and

\be
g(T) = \sum_{i=Bosons} g_i \left(\frac{T_i}{T_{\gamma}}\right)^3 +\frac{7}{8}\sum_{j=Fermions} g_j \left(\frac{T_j}{T_{\gamma}}\right)^3
\ee where $T_{i/j}$ are the temperatures of the individual relativistic species. Since the non thermal particles obey $sa^3=const$ the standard $g(T) a(T)^3 T_{\gamma}^3 = const$ still holds even in the presence of non-thermal species (assuming instantaneous reheating of the photon gas when species give off entropy upon annihilation), namely

\be
\frac{T_d(t)}{T_{\gamma}(t)} = \left(\frac{2}{g_d}\right)^{1/3} \rightarrow T_d(t) = \left(\frac{2}{g_d}\right)^{1/3} T_{\gamma,0} (1+z)
\ee where $T_d, g_d$ are the temperature and effective degrees of freedom at decoupling and $T_{\gamma,0}$ is the CMB temperature today.

Choosing the normalization $a_{today}=1$, the temperature evolves as $T(t) =T_0/a(t)$, where $T_0$ is the temperature of the plasma today ($T_0 = (2/g_d)^{1/3} T_{\gamma,0}$) we can rewrite the density and pressure by introducing the dimensionless quantities, $x=m/T(t), y=q_f(t)/T(t)=q_c/T_0$ to give

\be \label{energydensity}
\rho = \frac{g m}{2 \pi^2} T^3(t) \Bigg\langle y^2 \sqrt{1+\frac{y^2}{x^2}} \Bigg\rangle ~~;~~ \mathcal{P} = \frac{g}{6 \pi^2 m} T^5(t) \Bigg\langle \frac{y^4}{ \sqrt{1+\frac{y^2}{x^2}}} \Bigg\rangle ~~;~~ \langle g(x,y)\rangle \equiv \int dy \, g(x,y) f_d(y)
\ee where we've introduced the definition of $\langle g(x,y) \rangle$. Then the equation of state parameter is given by

\be \label{eos}
w = \frac{\mathcal{P}}{\rho} = \frac{1}{3 x^2} \frac{ \Bigg\langle \frac{y^4}{ \sqrt{1+\frac{y^2}{x^2}}} \Bigg\rangle}{\Big\langle y^2 \sqrt{1+\frac{y^2}{x^2}} \Big\rangle}
\ee  For non-relativistic species $x \gg 1$ so we neglect the $(y/x)^2$ terms to arrive at the familiar result

\bea
\rho_{nr} = m \frac{g }{2 \pi^2} T^3(t) \langle y^2  \rangle = m n(t) & ; & \mathcal{P}_{nr} = \frac{g}{6 \pi^2 m} T^5(t) \langle y^4 \rangle \nonumber \\
 w_{nr}  =  \frac{T(t)^2}{3 m^2} \frac{ \langle y^4 \rangle}{\langle y^2 \rangle} & \rightarrow & 0  \,.
\eea For relativistic species, $x \ll 1$ and $\mathcal{P}_{rel} =  \rho_{rel}/3$. Explicitly, the thermodynamic quantities become

\bea
\rho_{rel} = \frac{g }{2 \pi^2} T^4(t) \langle y^3  \rangle & ; & \mathcal{P}_{rel} = \frac{g}{6 \pi^2 } T^4(t) \langle y^3 \rangle \nonumber \\
w_{rel} = \frac{ \langle y^3 \rangle}{3 \langle y^3 \rangle} & = & \frac{1}{3} \,.
\eea

In the non relativistic limit, the average velocity per particle is given by
\be
\overline{ \vec{V}^2 } = \frac{\overline{\vQ^2}}{m^2} = \frac{T(t)^2}{m^2}\frac{\langle y^4 \rangle }{\langle y^2 \rangle} = \frac{3 \mathcal{P}}{\rho}
\ee which leads to the velocity dispersion relation

\be
\mathcal{P} = \sigma^2 \rho ~~;~~ \sigma = \sqrt{\frac{\overline{ \vec{V}^2}}{3}}  = \frac{T(t)}{m}\sqrt{\frac{\langle y^4 \rangle }{3 \langle y^2 \rangle}}
\ee The work of Tremaine and Gunn \cite{tremainegunn} and Lynden-Bell \cite{violentrelax} argued that the phase space density may only decrease as a galaxy evolves (violent relaxation and phase mixing). The phase space density is related to observationally accessible quantities (galactic velocity dispersion and density), and therefore the primordial phase space density can be used as an upper bound to place limits on dark matter parameters. For dwarf galaxies, these observations are summarized in \cite{destri} and the phase space density is given by

\be
\mathcal{D} = \frac{n(t)}{\overline{\vQ^2}^{\, 3/2}} \,.
\ee The phase space density is completely determined by \emph{moments} of the distribution function after freezeout. In terms of an arbitrary distribution function, this is given by

\be
\mathcal{D} = \frac{g}{2 \pi^2} \frac{\langle y^2 \rangle^{5/2} }{\langle y^4 \rangle^{3/2}}
\ee

During galactic evolution, the phase space decreases from its primordial value \cite{dalcantonhogan,violentrelax}. Eventually, today, the particles will be non relativistic and, for a non relativistic particle, we have that

\be
\mathcal{D}_{nr} = \frac{n}{\overline{ Q_f^2}^{\, 3/2}} = \frac{\rho}{m^4 \overline{ \vec{V}^2}^{\, 3/2}} = \frac{1}{3^{3/2} m^4} \frac{\rho}{\sigma^3}
\ee For a primordial phase space density, $\mathcal{D}_p$, imposing the bound $\mathcal{D}_{p} \ge \mathcal{D}_{nr}$ gives us the constraint

\be
\mathcal{D}_{p} \ge \frac{1}{3^{3/2} m_{\nu_s}^4} \left. \frac{\rho}{\sigma^3} \right|_{today}
\ee  where $\rho,\sigma$ are observationally accessible. For galaxies that are dominated mostly by dark matter, namely dwarf spheroidals, this can be used to place a limit on the dark matter mass and mixing angle.

Another observational quantity that would be relevant for a sterile neutrino dark matter candidate is the number of effective neutrino species, $N_{eff}$. The standard method of obtaining the number of neutrinos from cosmology involves measuring the number of effective relativistic species from the CMB. The sterile neutrinos we consider in this work decouple while they are still relativistic (at $\sim 10-15 \,MeV$) as discussed in section \ref{distfunction}. After sterile decoupling, all the normal standard model species continue to decay/annihilate and eventually only the active neutrinos, electrons, positrons, baryons and photons remain. Each time a species decouples, the entropy of the decoupled particles is swapped into the remaining relativistic species via entropy conservation shown in eq \ref{entropycons}. Because $sa^3=constant$ the standard relation that relates the temperature of ultrarelativisitic decoupled particles to the photon temperature follows:

\be
\frac{T_{\nu}^{active}(t)}{T_{\gamma}(t)} = \left(\frac{4}{11}\right)^{1/3} ~~;~~ \frac{T_{\nu_s}(t)}{T_{\gamma}(t)} = \left( \frac{2}{g_d} \right)^{1/3} \,.
\ee

After photon reheating the expression for the relativistic energy density becomes

\be
\rho_{rel} = \rho_{\gamma}\left(1+\frac{7}{8}\left(\frac{4}{11}\right)^{4/3} N_{eff,0} + \frac{\rho_{\nu_s}}{\rho_{\gamma}}\right)
\ee where $\rho$ is given by \ref{energydensity} and $N_{eff,0}=3.046$ is the standard result with only the active neutrinos \cite{neffboltz}. The CMB is formed when $T_{\gamma} \approx 1 eV$ and if the sterile neutrinos are still relativistic at this time they may contribute to $N_{eff}$. During matter domination prior to photon decoupling, a relativistic sterile neutrino has energy density given by

\be
\rho_{\nu_s} = \frac{g_s T_{\nu_s}^4(t)}{2 \pi^2} \langle y^3  \rangle \,.
\ee Using $\rho_{\gamma} = \frac{2 \pi^2}{30}T^4_{\gamma}$ we get that

\be
\frac{\rho_{\nu_s}}{\rho_{\gamma}} = g_s \frac{30}{4\pi^4} \left(\frac{T_{\nu_s}(t)}{T_{\gamma}(t)}\right)^4 \langle y^3 \rangle \,.
\ee So writing

\be
\rho_{rel} = \rho_{\gamma}\left(1+\frac{7}{8}\left(\frac{4}{11}\right)^{4/3} \Big(N_{eff,0}+\Delta N_{eff}\Big)\right)
\ee leads to the defintion

\be \label{neff}
\Delta N_{eff} = \left(\frac{11}{4} \frac{2}{g_d}\right)^{4/3}\frac{60 g_{\nu_s}}{7\pi^4} \langle y^3 \rangle
\ee where $N_{eff} = N_{eff,0} + \Delta N_{eff}$ has been most recently measured by the Planck satellite \cite{planck}. The results above are general and all that remains is to obtain $f_d(y)$ for a particular mechanism.

\section{Quantum Kinetic Equation}

It is generally accepted that in the early universe, where temperatures and densities are larger than the QCD scale ($\sim 155 MeV$), quarks and gluons are asymptotically free forming a quark-gluon plasma. As the universe expands and cools, quarks and gluons undergo two phase transitions: deconfinement/confinement and chiral symmetry breaking.  Confinement and hadronization result predominantly in the formation of baryons and mesons, the lightest of which - the pions - are dominant and are the pseudo Goldsone bosons associated with chiral symmetry breaking \cite{schwarz}. A recent lattice QCD calculation \cite{hotqcd} suggests that this phase transition is not first order but a rapid crossover near a critical temperature $T_{QCD} \approx 155 MeV$. Pions thermalize in the plasma via strong, electromagnetic and weak interactions and are in local thermodynamic equilibrium. Their decay into leptons and active neutrinos is balanced by the inverse process as the leptons and active neutrinos are also in LTE. However if the pions (slowly) decay into sterile neutrinos, detailed balance will not be maintained as the latter are not expected to be in LTE.

As the pion is the lowest lying bound state of QCD, it is a reasonable assumption that during the QCD phase transition pions will be the most dominantly produced bound state. During this time, pions will remain in LTE with the \emph{active} neutrinos by detailed balance $\pi \rightleftharpoons l \nu_{a}$. We focus on sterile neutrino $\nu_s$ production from $\pi \rightarrow l \nu_s$ which is suppressed by $|U_{ls}|^2 \ll 1$ with respect to the active neutrinos and does not maintain detailed balance. We also restrict the analysis to a scenario with no lepton asymmetry which sets the chemical potential of pions and leptons to zero. The interaction Hamiltonian responsible for this decay is

\be
H_i = \sum_{l=e,\mu} \sqrt{2} G_F V_{ud} f_{\pi} \int d^3x \left[ \bar{\Psi}_{\nu_{l}}(x,t) \gamma^{\sigma} \mathbbm{L} \Psi_{l} (x,t) J^{\pi}_{\sigma} (\vx,t) + H.C. \right]
\ee where $ J^{\pi}_{\sigma} = i \partial_{\sigma} \pi (x,t)$ is the pseudoscalar pion current.

\begin{figure}[h!]
\includegraphics[height=4.0in,width=4.0in,keepaspectratio=true]{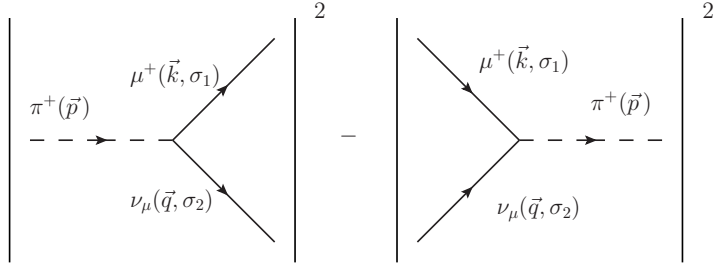}
\caption{The gain/loss terms for the quantum kinetic equation describing $\pi^{+} \rightarrow \bar{\mu} \nu_{\mu}$.}
\label{fig:kinetic}
\end{figure}

The buildup of the daughter particles can be described via a quantum kinetic equation that takes the usual form of

\be
\frac{dn}{dt}(q,t) = \delta n_{Gain} - \delta n_{Loss} = \mathbbm{P}[n(t)]
\ee where the gain and loss terms are obtained from the appropriate transition probabilities $|\mathcal{M}_{fi}|^2$. For this Hamiltonian, the processes relevant for neutrino build up are displayed in fig \ref{fig:kinetic}.

 The gain terms arise from the reaction $\pi^{+} \rightarrow \bar{l} \nu_{l}$ where the initial state has $N_{\vp}$ quanta of pions and $n_{\vk,s'},n_{\vq,s}$ quanta of charged leptons and neutrinos respectively while the final state has quanta $N_{\vp}-1,n_{\vk,s'}+1,n_{\vq,s}+1$ for the respective species. The Fock states for the gain process are given by

\be
|i \rangle = | N^{\pi^+}_{p}, n^{\bar{l}}_{k}, n^{\nu}_{q} \rangle ~~;~~ |f \rangle = | N^{\pi^+}_{p} - 1, n^{\bar{l}}_{k}+1, n^{\nu}_{q} +1 \rangle .
\ee Similarly, the loss terms are obtained from the reverse reaction $\bar{l} \nu_{l} \rightarrow \pi^+$ where the initial state has $N_{\vp},n_{\vk,s'},n_{\vq,s}$ quanta for pions, charged leptons and neutrinos respectively. The final state has $N_{\vp}+1,n_{\vk,s'}-1,n_{\vq,s}-1$ of the appropriate quanta and the Fock states for the loss process are given by

\be
|i \rangle = | N^{\pi^+}_{p}, n^{\bar{l}}_{k}, n^{\nu}_{q} \rangle ~~;~~ |f \rangle = | N^{\pi^+}_{p} + 1, n^{\bar{l}}_{k}-1, n^{\nu}_{q} - 1 \rangle .
\ee The neutrino flavor states are expanded in terms of the mass eigenstates via the $U_{PMNS}$ matrix as per usual

\be
|\nu_{\alpha} \rangle = \sum_i U^*_{\alpha i} | \nu_i \rangle
\ee

Through a standard textbook calculation, the transition amplitudes at first order in perturbation theory can be calculated. The matrix element relevant for the gain term is given by

\bea
 \mathcal{M}_{fi} |_{gain} & = & -i\int d^4x \langle N^{\pi^+}_{p} - 1, n^{\bar{l}}_{k}+1, n^{\nu}_{q} +1 |  \mathcal{H}_I(x) | N^{\pi^+}_{p}, n^{\bar{l}}_{k}, n^{\nu}_{q} \rangle \\
\nonumber & = & i \sqrt{2} G_F V_{ud} f_{\pi} \sum_i U^*_{l i} \frac{2 \pi}{\sqrt{V}} \frac{ \bar{\mathcal{U}}^{\nu_i}(q,\sigma_1) \slashed{p} \mathbbm{L} \mathcal{V}^{l}(k,\sigma_2) }{\sqrt{8 E_{\pi}(p) E_{l}(k) E_{\nu}(q) } } \\
\nonumber & * & \delta_{\vp,\vk+\vq} \, \delta(E_{\pi}(p) - E_{\nu}(q) - E_{l}(k) \sqrt{N_{\pi}(p)} \sqrt{1-n_{l}(k)} \sqrt{1-n_{\nu}(q)} \,
\eea and the matrix element relevant for the loss term is given by

\bea
 \mathcal{M}_{fi} |_{loss} & = & -i\int d^4x \langle N^{\pi^+}_{p} + 1, n^{\bar{l}}_{k}-1, n^{\nu}_{q} - 1 | \mathcal{H}_I(x) |N^{\pi^+}_{p}, n^{\bar{l}}_{k}, n^{\nu}_{q}\rangle \\
 \nonumber & = & i \sqrt{2} G_F V_{ud} f_{\pi} \sum_i U_{l i} \frac{2 \pi}{\sqrt{V}} \frac{ \bar{\mathcal{V}}^{\, l}(k,\sigma_2) \slashed{p} \mathbbm{L} \mathcal{U}^{\nu_i}(q,\sigma_1) }{\sqrt{8 E_{\pi}(p) E_{l}(k) E_{\nu}(q) } }\\
 \nonumber & * & \delta_{\vp,\vk+\vq \,} \delta(E_{\pi}(p) - E_{\nu}(q) - E_{l}(k)) \sqrt{N_{\pi}(p)+1} \sqrt{n_{l}(k)} \sqrt{n_{\nu}(q)}
\eea Restricting our attention towards the production of one particular mass eigenstate, $i=s$ (ie $\pi \rightarrow \bar{\mu} \nu_i$ as opposed to $\pi \rightarrow \bar{\mu} \nu_{\mu}$) will give the production distribution of a sterile neutrino. The idea is that the active neutrinos will remain in thermal and chemical equilibrium through $\pi \rightleftarrows \bar{l} \nu_l$ but if we assume that there had been \emph{no sterile neutrino production} prior to pion decays then this will be the dominant contribution to sterile neutrino population. With this adjustment, summing over $\vk,\vp,\sigma_1,\sigma_2$ leads to the production rate

\bea
 \frac{1}{T} \sum_{\vk,\vp,\sigma_1,\sigma_2} |\mathcal{M}_{fi} |^2_{gain} & = & |U_{l s}|^2 |V_{u d}|^2 \frac{\pi G_F^2 f_{\pi}^2}{2} \int \frac{d^3 p}{(2\pi)^3} \frac{N_{\pi}(p) (1-n_{l}(k))(1-n_{\nu}(q))}{E_{\pi}(p) E_{\mu}(k) E_{l}(q)} \\
 \nonumber & * & Tr[\slashed{p} \mathbbm{L} (\slashed{q} + m_s ) \slashed{p} \mathbbm{L} (\slashed{k} - m_{l}) ] \delta_{\vp,\vk+\vq \,} \delta(E_{\pi}(p) - E_{\nu}(q) - E_{l}(k)) \\
 \nonumber & = & \frac{|U_{l s}|^2 |V_{u d}|^2 G_F^2 f_{\pi}^2}{8 \pi^2} \int \frac{d^3 p}{(2\pi)^3} \frac{N_{\pi}(p) (1- n_{l}(k))(1-n_{\nu}(q))}{E_{\pi}(p) E_{l}(k) E_{\nu}(q)} \\
 \nonumber & * & [2 (p \cdot q) (p \cdot k) - p^2 (q \cdot k) ]   \delta_{\vp,\vk+\vq \,} \delta(E_{\pi}(p) - E_{\nu}(q) - E_{l}(k))
\eea where T stands for the total interaction time, not to be confused with temperature, and the evaluation of the matrix elements is a standard exercise.  The loss term is calculated in the same way but with the substitution $N \rightarrow 1 + N$ and $1-n \rightarrow n$. With the aforementioned replacements and using the energy/momentum conserving delta functions leads to the rate equation

\bea \label{rateeqn}
\frac{dn}{dt} & = & \frac{1}{T} \sum_{\vk,\vp,\sigma_1,\sigma_2} |\mathcal{M}_{fi} |^2_{gain} -  |\mathcal{M}_{fi} |^2_{loss} \\ \nonumber & = & \frac{ |U_{l s} |^2 |V_{u d}|^2 G_F^2 f_{\pi}^2 }{8 \pi} \frac{m^2_{\pi}(m^2_{l} + m^2_{\nu} ) - (m^2_{l} -m^2_{\nu})^2}{q E_{\nu}(q)}  \\
 \nonumber & * & \int^{p_+}_{p_-} \frac{dp \, p}{\sqrt{p^2 + m_{\pi}^2 } } \Big[ N_{\pi}(p) (1 - n_{\bar{l}}(\vp-\vq \,) )(1 - n_{\nu}(q)) - (1+N_{\pi}(p)) n_{\bar{l}}(\vp-\vq \,) n_{\nu}(q) \Big]
\eea where $p_{\pm}$ are obtained from the constraint

\be
[ (|\vp|-|\vq|)^2 + m^2_{l} ]^{1/2} \leq E_{\pi}(p) - E_{\nu}(q) \leq [ (|\vp|+|\vq|)^2 + m^2_{l} ]^{1/2}  .
\ee This gives the solutions

\be
p_{\pm} = \left| \frac{E_{\nu}(q)}{2m_{\nu}^2} [ (m_{\pi}^2 - m^2_{l} + m_{\nu}^2)^2 -4m^2_{\pi}m^2_{\nu} ]^{1/2} \pm \frac{q(m^2_{\pi} - m^2_{l} + m^2_{\nu} )}{2m^2_{\nu}} \right|  \,. \label{limits}
\ee Note that these bounds coalesce at when $m_{\pi}^2-m_l^2+m_{\nu}^2 = 2 m_{\pi}^2 m_{\nu}^2$ and the rate, Eq \ref{rateeqn}, vanishes simply because this corresponds to the reaction's kinematic threshold. These results are extended to the early universe by replacing the momentum with the physical momentum, $q \rightarrow Q_f = q_c/a(t)$, and use of the results from section \ref{decoupleddynamics}.

\section{Non-thermal Sterile Neutrino Distribution Function} \label{distfunction}

A body of work \cite{tytgat1,nicola1,nicola2,jeon,harada} has established that, when $\pi$'s are present in the medium in LTE, the $\pi$ decay constant, $f_{\pi}$, and $\pi$ mass vary with temperature for $T \lesssim T_{QCD}$ where $T_{QCD}$ is the critical temperature for the QCD phase transition. We account for these effects and make several simplifications by implementing the following:

\begin{itemize}
\item The finite-temperature pion decay constant has been obtained in both non-linear sigma models \cite{tytgat1} and Chiral perturbation theory \cite{nicola1,nicola2,jeon,harada} with the result given as

\be
f^2_{\pi} \rightarrow f^2_{\pi}(t) = f^2_{\pi}(0)\left(1-\frac{T(t)^2}{6 f_{\pi}(0)^2}\right) ~~;~~ f_{\pi}(0) = 93 MeV\,.
\ee This result is required in the quantum kinetic equation since production begins near $T_{QCD} \sim 155 MeV$ and continues until the distribution function freezes out. We assume prior to $T_{QCD}$ that there are no pions and that hadronization happens instantaneously at $T \sim T_c \sim 155 MeV$.

\item The mass of the pion varies with temperature as described in detail in ref \cite{nicola1,nicola2}. The finite temperature corrections to the pion mass is calculated with electromagnetic corrections in chiral perturbation theory and its variation with temperature is shown in figure 2 of \cite{nicola1}. In these references it can be seen that between 50 and 150 MeV the pion mass only varies between 140 and 144 MeV. Since this change is so small, we neglect the temperature variation in the pion mass and simply use its average value: $m_{\pi}(T) = 142 MeV$ (see fig in ref \cite{nicola1}).

\item We assume that the lepton asymmetry in the early universe is very small so that we may neglect the chemical potential in the distribution function of the pions and charged leptons. This asymmetry is required for the Shi-Fuller mechanism but will not be present in these calculations. We will show a similar enhancement at low moment to SF but the enhancement found here is with zero lepton asymmetry.

\item With the assumption that there is no lepton asymmetry, the contribution to thermodynamic quantities from $\pi^{-} \rightarrow l \bar{\nu}$ will be equal to that of $\pi^{+} \rightarrow \bar{l} \nu$. In which case, the degrees of freedom will be set at $g_{\nu} = 2$ accounting for both equal particle and antiparticle contributions in the case of Dirac fermions and the two different sources ($\pi^{\pm}$) for Majorana fermions. The different helicities have already been accounted by summing over spins in the evaluation of the matrix elements of the previous section.

\item We assume that there had been no production of sterile neutrinos prior to the hadronization period from any other mechanisms (such as scalar decays or DW). This allows us to set the initial distribution of the sterile neutrinos to zero in the kinetic equation which implies that our results for the distribution function will be a \emph{lower bound for the distribution function}. Any other prior sources could only enhance the population of sterile neutrinos. By neglecting the initial population, we can neglect the Pauli blocking factor of the $\nu$'s in the production term and we can also neglect the loss term (see discussion below).

\end{itemize}

 After the QCD phase transition, there is an abundance of pions present in the plasma in thermal/chemical equilibrium. The pions will decay, predominantly via $\pi^{\pm} \rightarrow l^{\pm} \nu_s \, (\bar{\nu}_s) $, producing sterile neutrinos which, assuming that sterile neutrinos had not been produced up to this point, will have a negligible distribution function.  The reverse reaction ($\bar{l} \nu_s \rightarrow \pi$) will not occur in any significant quantities also due to the assumption of null initial population and $|U_{ls}|^2 \ll 1$; under these assumptions we may neglect the loss terms in the kinetic equation. With these assumptions, we use the following distributions for the production terms in the quantum kinetic equation

\be
N_{\pi} = \frac{1}{e^{E_{\pi}(p,t)/t} - 1} ~~;~~ n_{l} = \frac{1}{e^{E_{l}(p,t)/t} + 1} ~~;~~ n_{\nu_{s}} \approx 0 ~~;~~ E_{\alpha}(k,t) = \sqrt{\frac{k_c^2}{a(t)^2} + m_{\alpha}^2}
\ee where $k_c$ is a comoving momentum as discussed in section \ref{decoupleddynamics}.

With these replacements, neglecting the loss terms and setting $E_{l}(p,q) = E_{\pi}(p) -E_{\nu}(q)$ the quantum kinetic equation becomes

\bea
\frac{dn}{dt}(q,t) & = & \frac{ |U_{l s} |^2 f_{\pi}(t)^2 }{16 \pi} \frac{m^2_{\pi}(m^2_{l} + m^2_{\nu} ) - (m^2_{l} -m^2_{\nu})^2}{q\sqrt{q^2+m_{\nu}^2}}  \\
\nonumber & * & \int^{p_+}_{p_-} \frac{dp \, p}{\sqrt{p^2 + m_{\pi}^2 } } \left[ \frac{e^{-E_{\nu}(q)/T} e^{E_{\pi}(p)/T}}{(e^{E_{\pi}(p)/T}-1) (e^{-E_{\nu}(q)/T} e^{E_{\pi}(p)/T}+1)} \right]
\eea where the limits of integration are given by Eq. \ref{limits} and we have suppressed the dependence of physical momentum on time. The integral can be done by a simple substitution with the final result given here

\bea \label{rate}
\frac{dn}{dt}(q,t) & = & \frac{ |U_{l s} |^2 f_{\pi}^2(t) }{16 \pi } \frac{m^2_{\pi}(m^2_{l} + m^2_{\nu} ) - (m^2_{l} -m^2_{\nu})^2}{q(t) E_{\nu}(q,t) (e^{E_{\nu}(q,t)/T(t)}+1)} T(t) \\
 \nonumber & * & \ln \left( \frac{  1-e^{-\sqrt{p^2+m_{\pi}^2}/T(t)}  }{  e^{-E_{\nu}(q,t)/T(t)} + e^{-\sqrt{p^2+m_{\pi}^2}/T(t)}   } \right) \Bigg|^{p=p_+(t)}_{p=p_-(t)}
\eea where $p^{\pm}$ are given by Eq. \ref{limits}.

We make the following change of variables

\be
\tau = \frac{m_{\pi}}{T(t)} ~~;~~ \frac{d\tau}{dt} = \tau H(t) ~~;~~ y = \frac{p(t)}{T(t)} = \frac{p_c}{T_0}
\ee where $T_0$ is the temperature of the plasma \emph{today} since the normalization is set by $a(t_0)=1$. The QCD phase transition begins deep inside the radiation dominated epoch as does freeze out (see below) so that the Hubble factor is given by eq \ref{hubble}. Inserting the form of the Hubble factor into eq \ref{rate} prompts the convenient definition

\be
\Lambda = \frac{|U_{ls}|^2}{\sqrt{g(t)}} \left[ \frac{ |V_{u d}|^2 f^2_{\pi}(0) G^2_F}{8 \pi*1.66} \frac{M_{pl}}{m_{\pi}} \right] \left(m_l^2 +m_{\nu_s}^2 -\frac{(m_l^2-m_{\nu_s}^2)^2}{m_{\pi}^2} \right) \,.
\ee During the period shortly after hadronization when $m_{\mu} \lesssim T \lesssim m_{\pi}$ the relativistic degrees of freedom are $g(t) \sim  14.25$ while in the regime $m_{e} \lesssim T \lesssim m_{\mu}$ the degrees of freedom count is $g(t) \sim 10.75$ \cite{pdg}. We expect the sterile neutrino decoupling to happen well above the electron mass (this will be justified later) and since the variation of $g(t)$ is small we replace it with its average value, $g(t) \sim \bar{g} = 12.5$, so that we can neglect the time dependence of $\Lambda$.

These substitutions and variable changes lead to a more tractable form of the kinetic equation

\be \label{exactrate}
\frac{1}{\Lambda}\frac{dn}{d\tau}(y,\tau) = \frac{(\tau/y)^2 (1- \frac{m_{\pi}^2/6 f^2_\pi}{\tau^2})}{\sqrt{1+\frac{m_{\nu_s}^2}{m_{\pi}^2}\frac{\tau^2}{y^2}}\left( e^{E^q_{\nu}/T}+1\right) } \ln \left( \frac{  1-e^{-\sqrt{p^2+m_{\pi}^2}/T(t)}  }{  e^{-E_{\nu}(q,t)/T(t)} + e^{-\sqrt{p^2+m_{\pi}^2}/T(t)}   } \right) \Bigg|^{p=p_+(t)}_{p=p_-(t)} \,.
\ee The population build up is obtained by integrating

\be
n(\tau,y) = \int^{\tau}_{\tau_0} d\tau' \frac{dn}{d\tau}(\tau',y) \,.
\ee where we have neglected any early population of $\nu_s$ and the value of $\tau_0$ is determined by when the pions are produced, assumed almost immediately after the hadronization transition. Our assumption is that this happens instantaneous at the QCD phase transition and the pions reach equilibrium instantaneously. This is justified by the results of \cite{hotqcd} which suggest a continuous transition which allows for thermalization on strong interaction time scales.

As shown in \cite{hotqcd}, the continuous phase transition occurs at $T_{QCD} \sim 155 MeV$ so that $\tau_0 = m_{\pi}/T_{QCD} = 0.92 \approx 1$. As we set $\tau_0$ below this value we expect that the population would increase as there will simply be more time for production to occur; this will be confirmed in a subsequent section.

\begin{figure}[h!]
\begin{center}
\includegraphics[height=3.5in,width=3.2in,keepaspectratio=true]{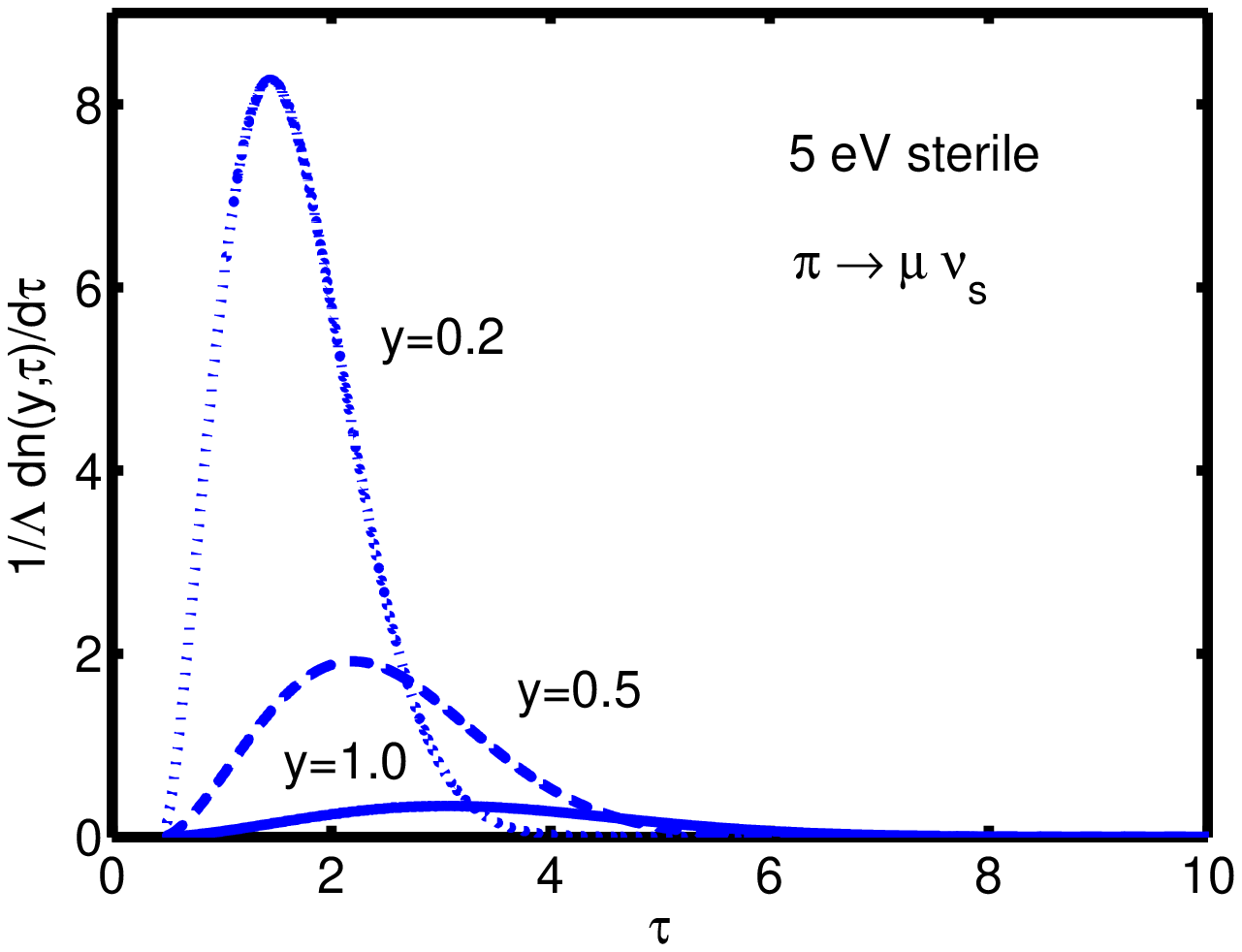}
\includegraphics[height=3.5in,width=3.2in,keepaspectratio=true]{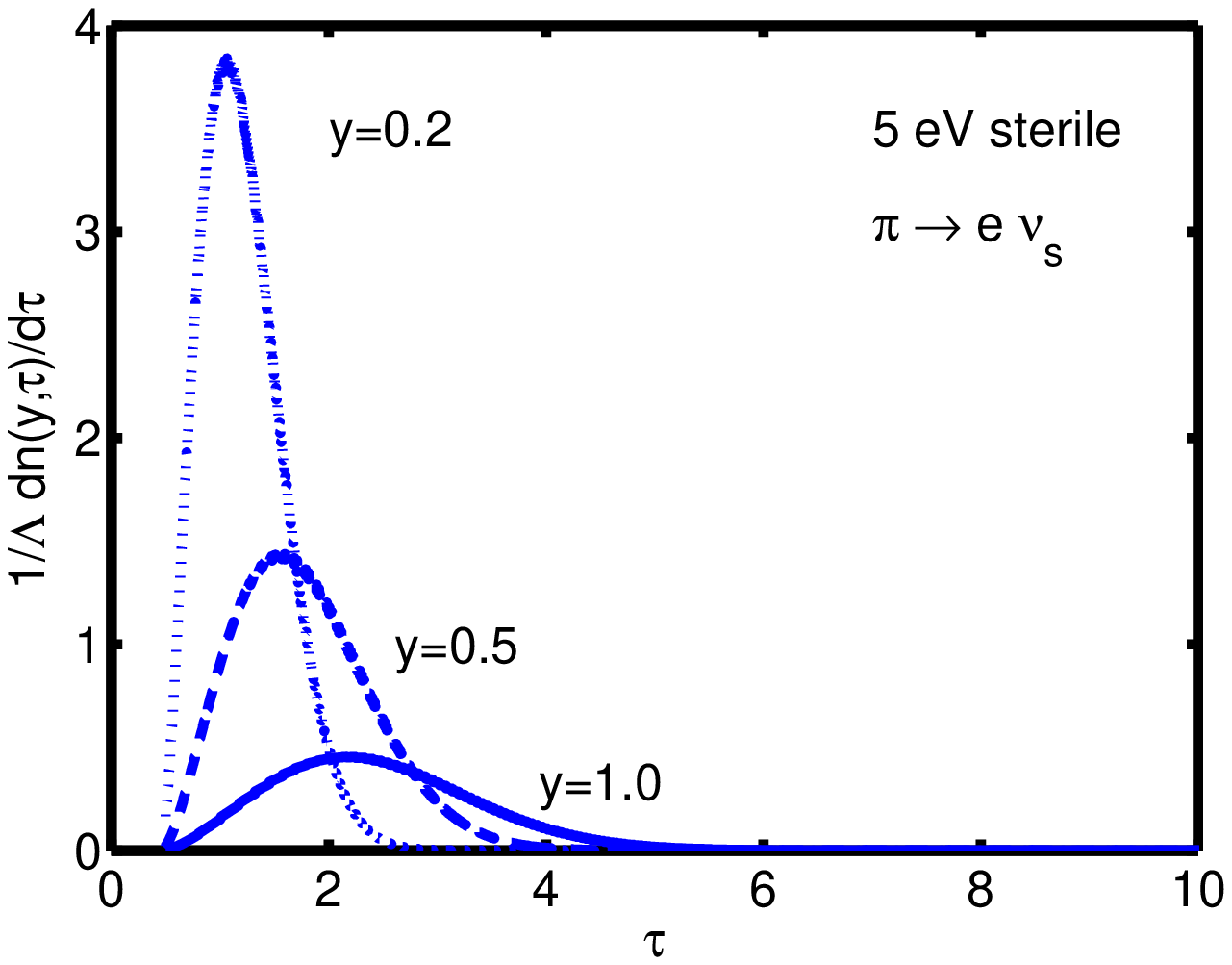}
\includegraphics[height=3.5in,width=3.2in,keepaspectratio=true]{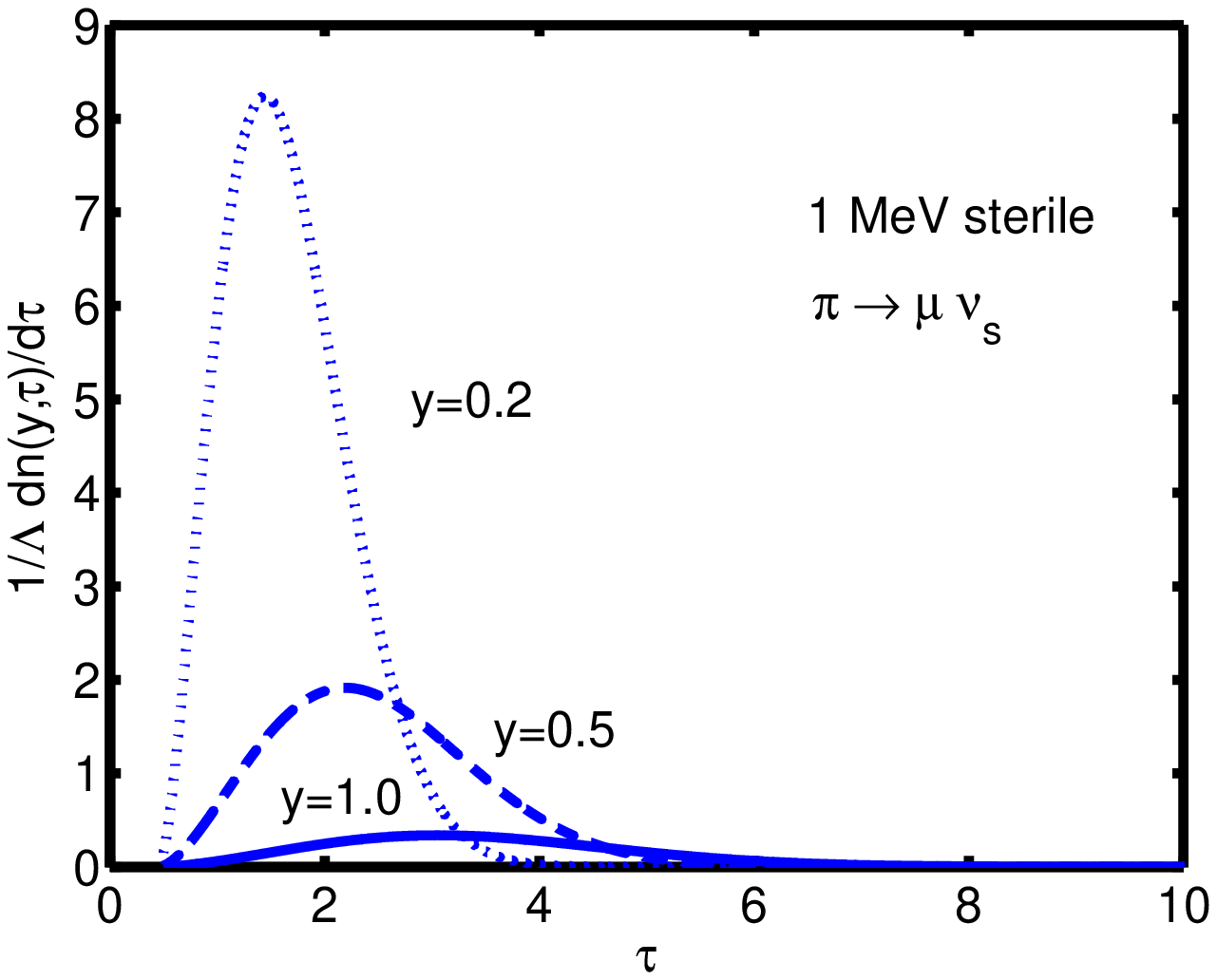}
\includegraphics[height=3.5in,width=3.2in,keepaspectratio=true]{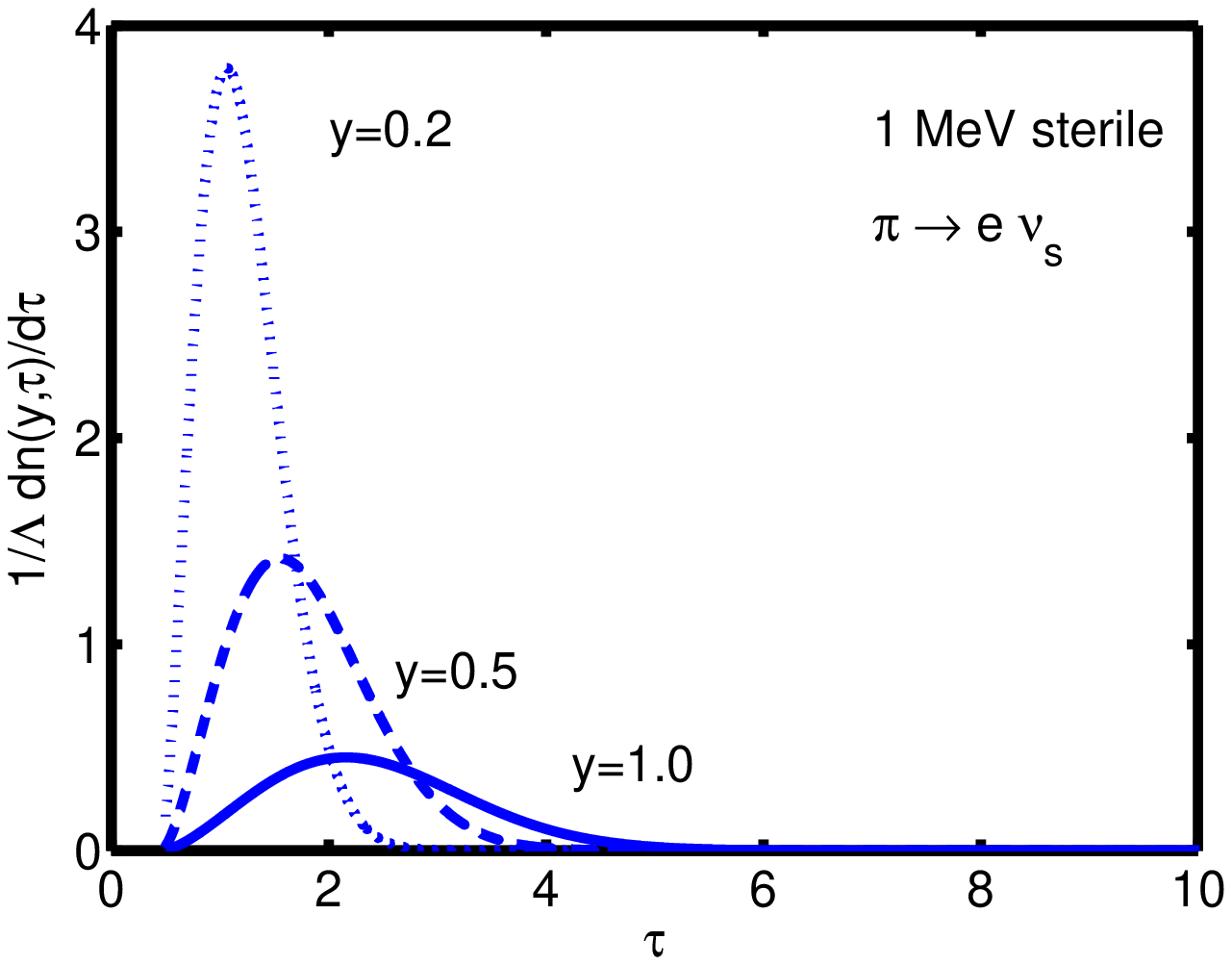}
\caption{Production rate of a sterile $\nu_s$ obtained from quantum kinetics from $\pi \rightarrow l \nu_s$ with $l=\mu,e$. Note that for $m_{\nu} \lesssim 1 MeV$ the rate does not vary significantly.}
\label{fig:evkinetics}
\end{center}
\end{figure}

\begin{figure}[h!]
\begin{center}
\includegraphics[height=3.5in,width=3.2in,keepaspectratio=true]{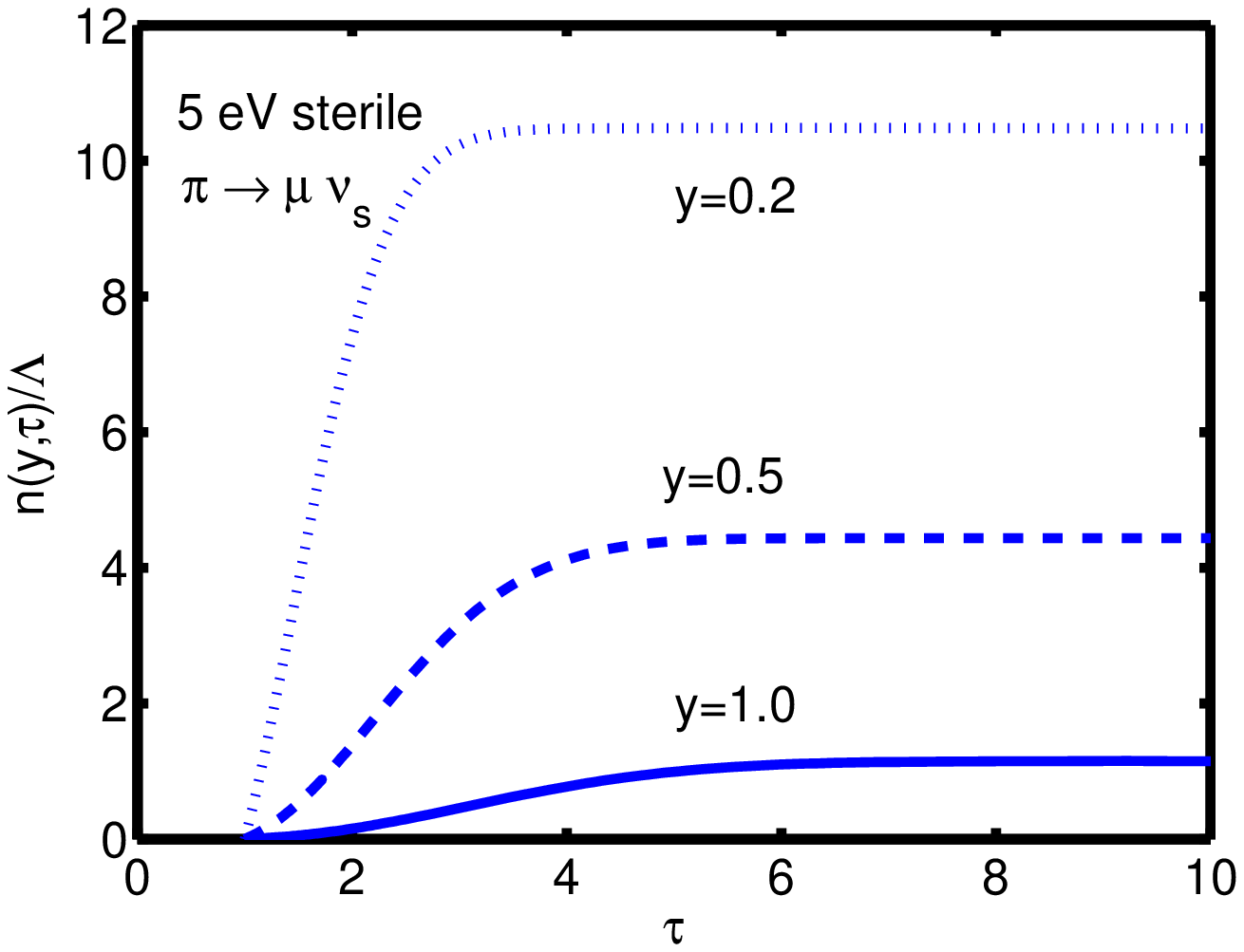}
\includegraphics[height=3.5in,width=3.2in,keepaspectratio=true]{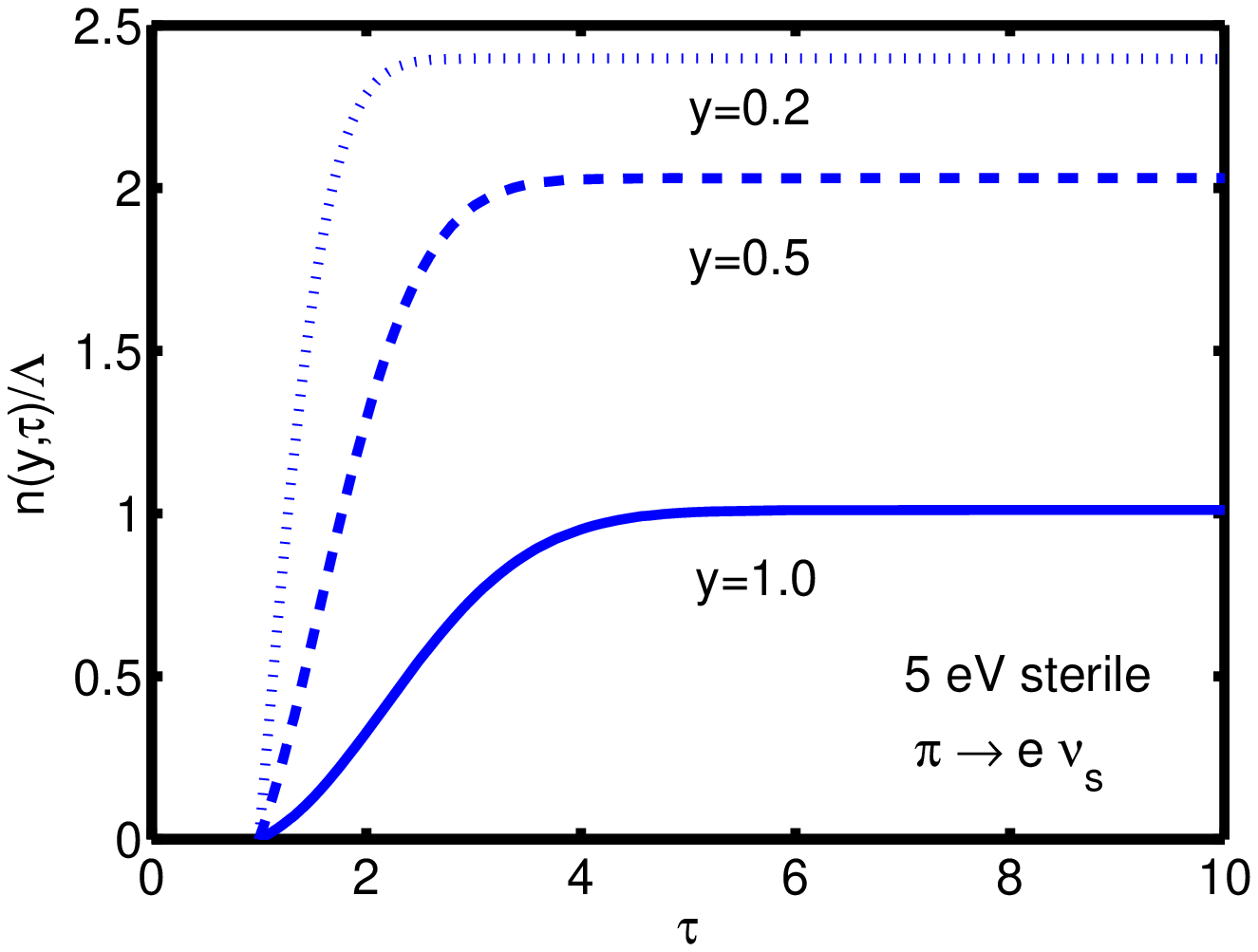}
\includegraphics[height=3.5in,width=3.2in,keepaspectratio=true]{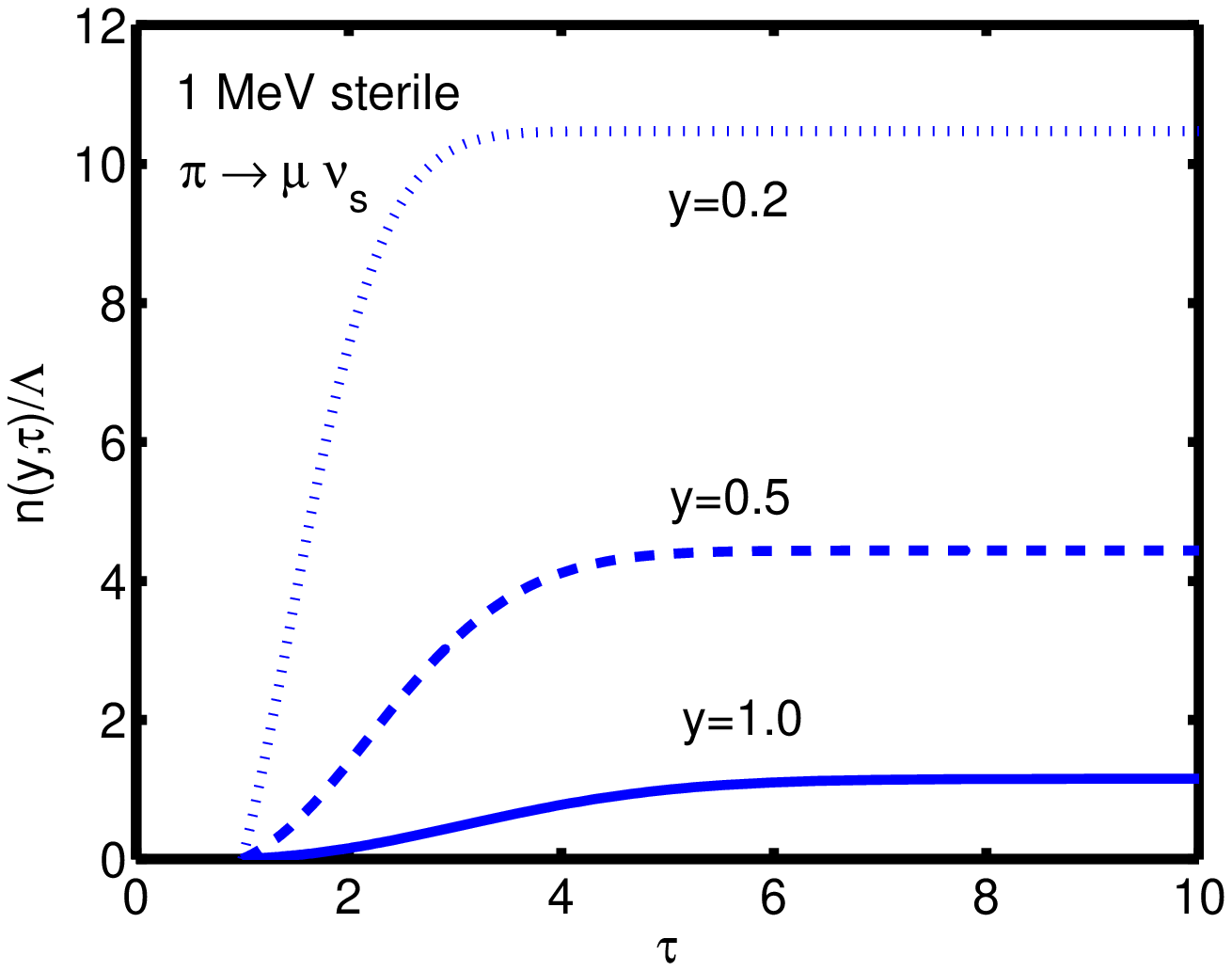}
\includegraphics[height=3.5in,width=3.2in,keepaspectratio=true]{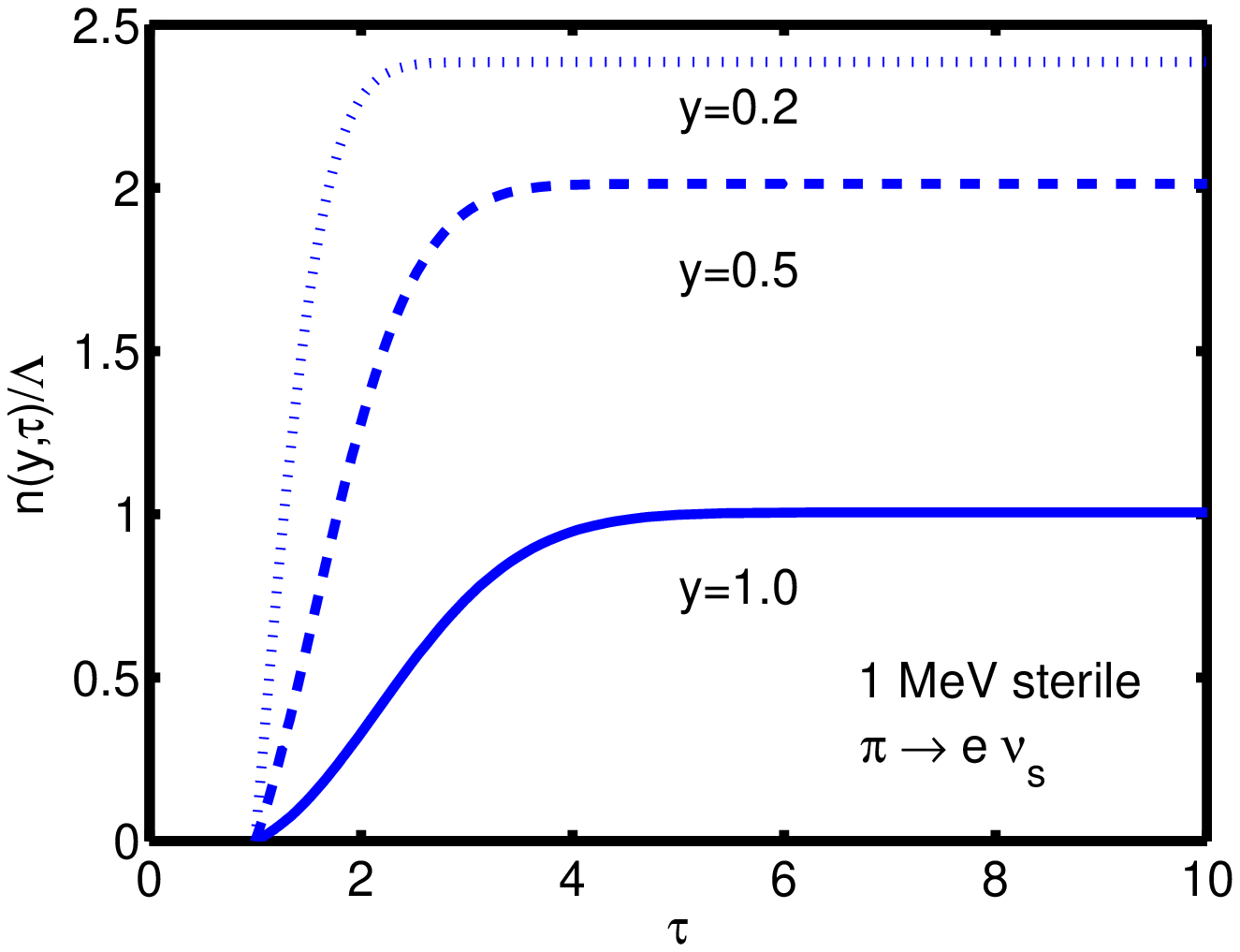}
\caption{Population build up of a sterile $\nu_s$ obtained from quantum kinetics from $\pi \rightarrow l \nu_s$ with $l=\mu,e$. Note that for $m_{\nu} \lesssim 1 MeV$ the build up does not vary significantly.}
\label{fig:mevkinetics}
\end{center}
\end{figure}

\begin{figure}[h!]
\begin{center}
\includegraphics[height=3.5in,width=3.2in,keepaspectratio=true]{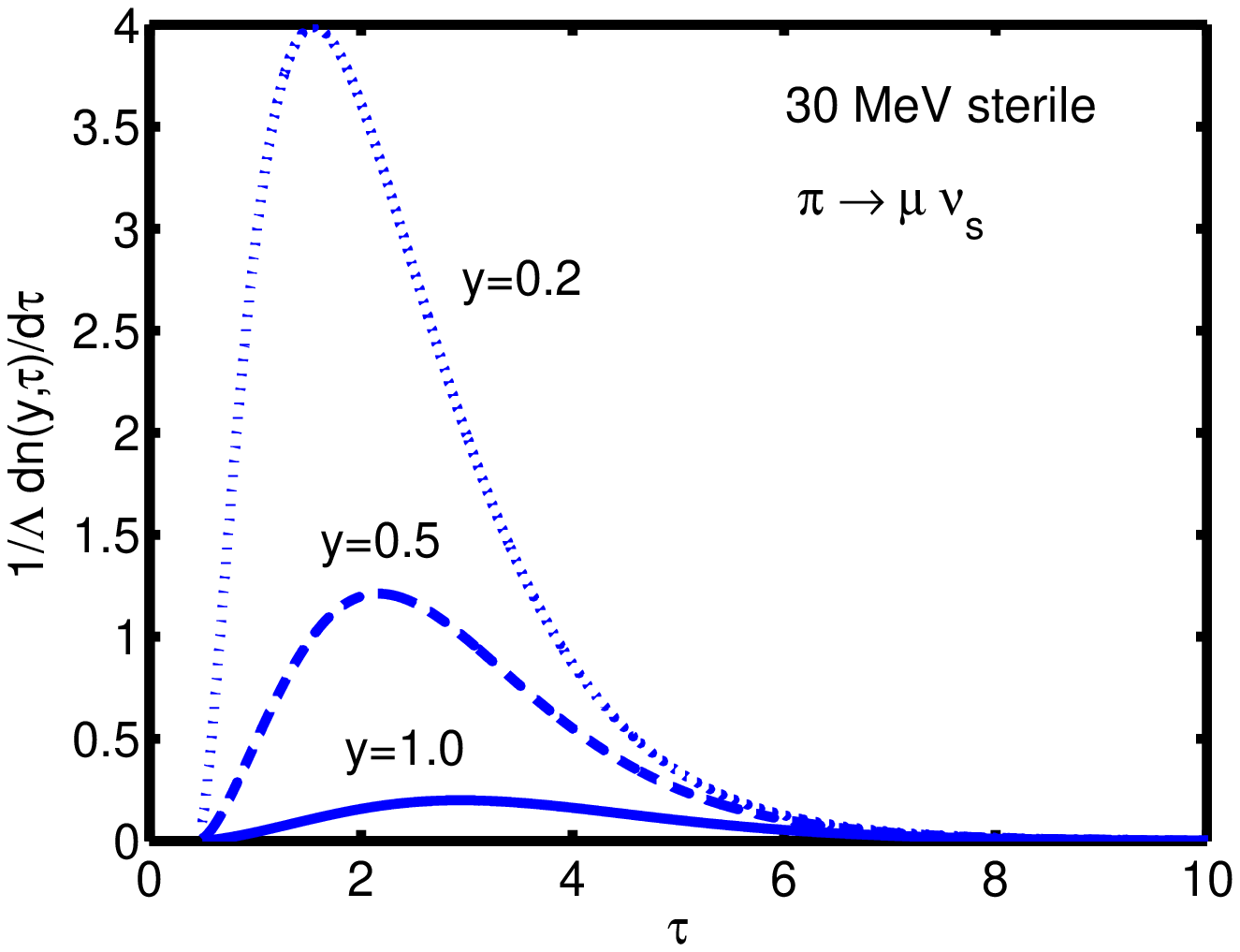}
\includegraphics[height=3.5in,width=3.2in,keepaspectratio=true]{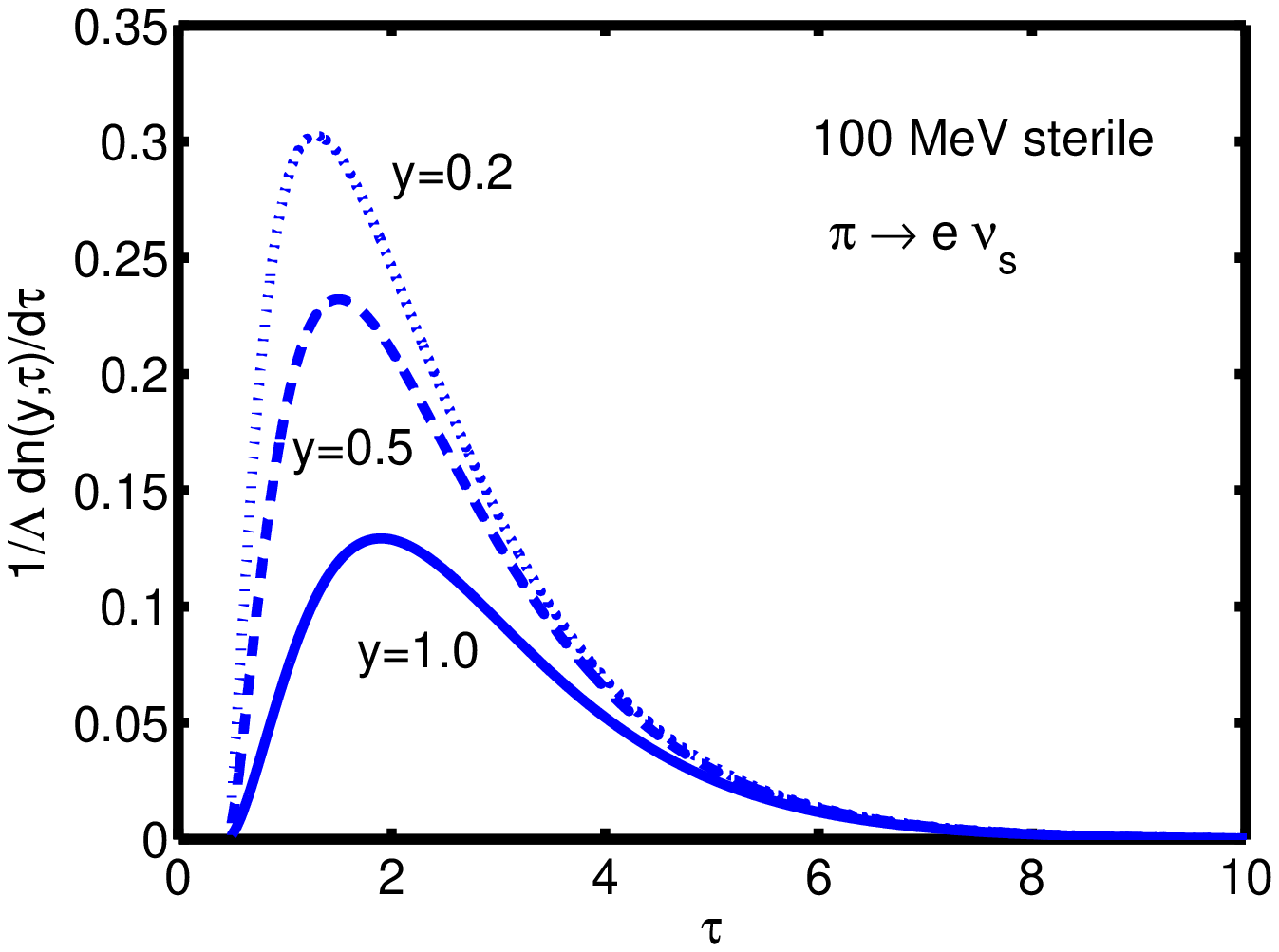}
\includegraphics[height=3.5in,width=3.2in,keepaspectratio=true]{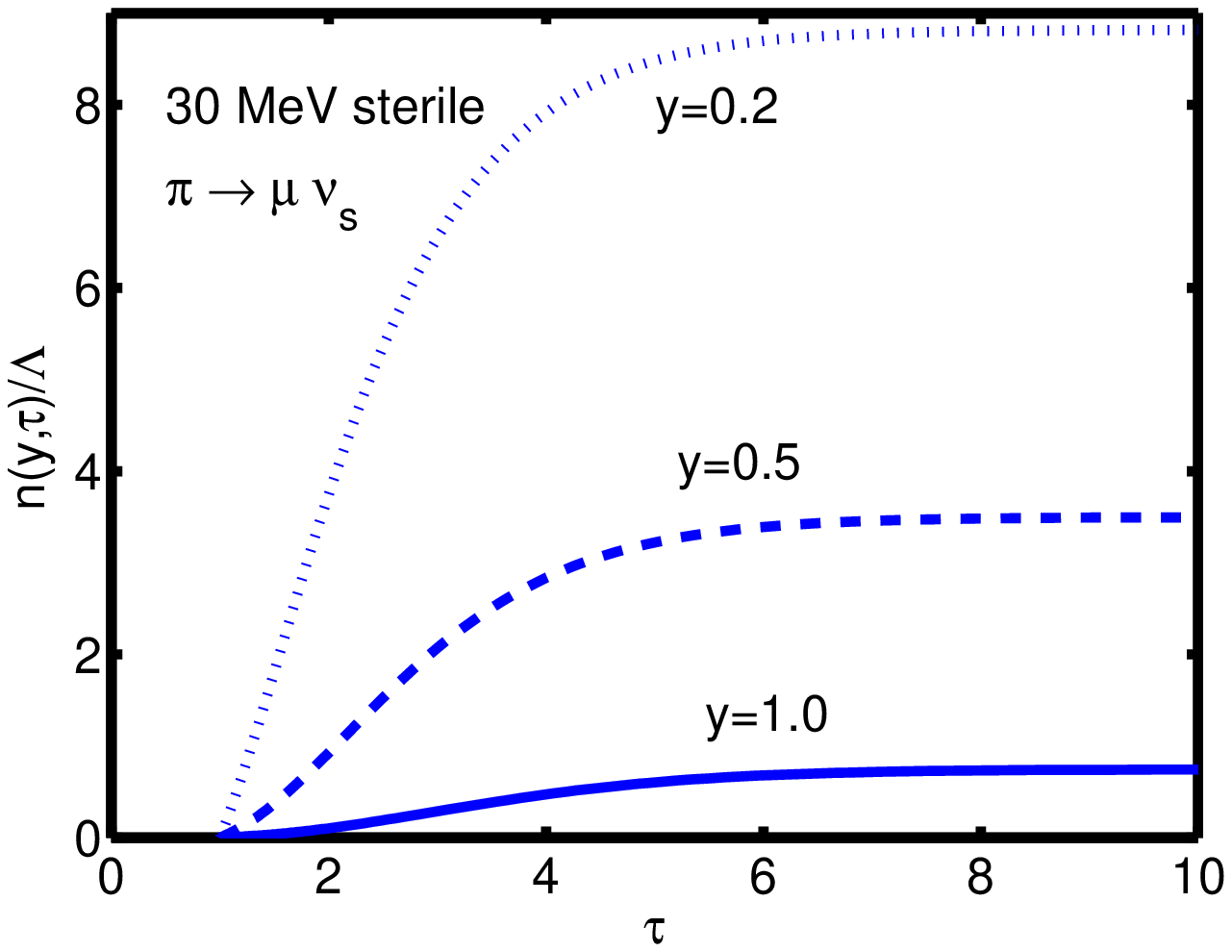}
\includegraphics[height=3.5in,width=3.2in,keepaspectratio=true]{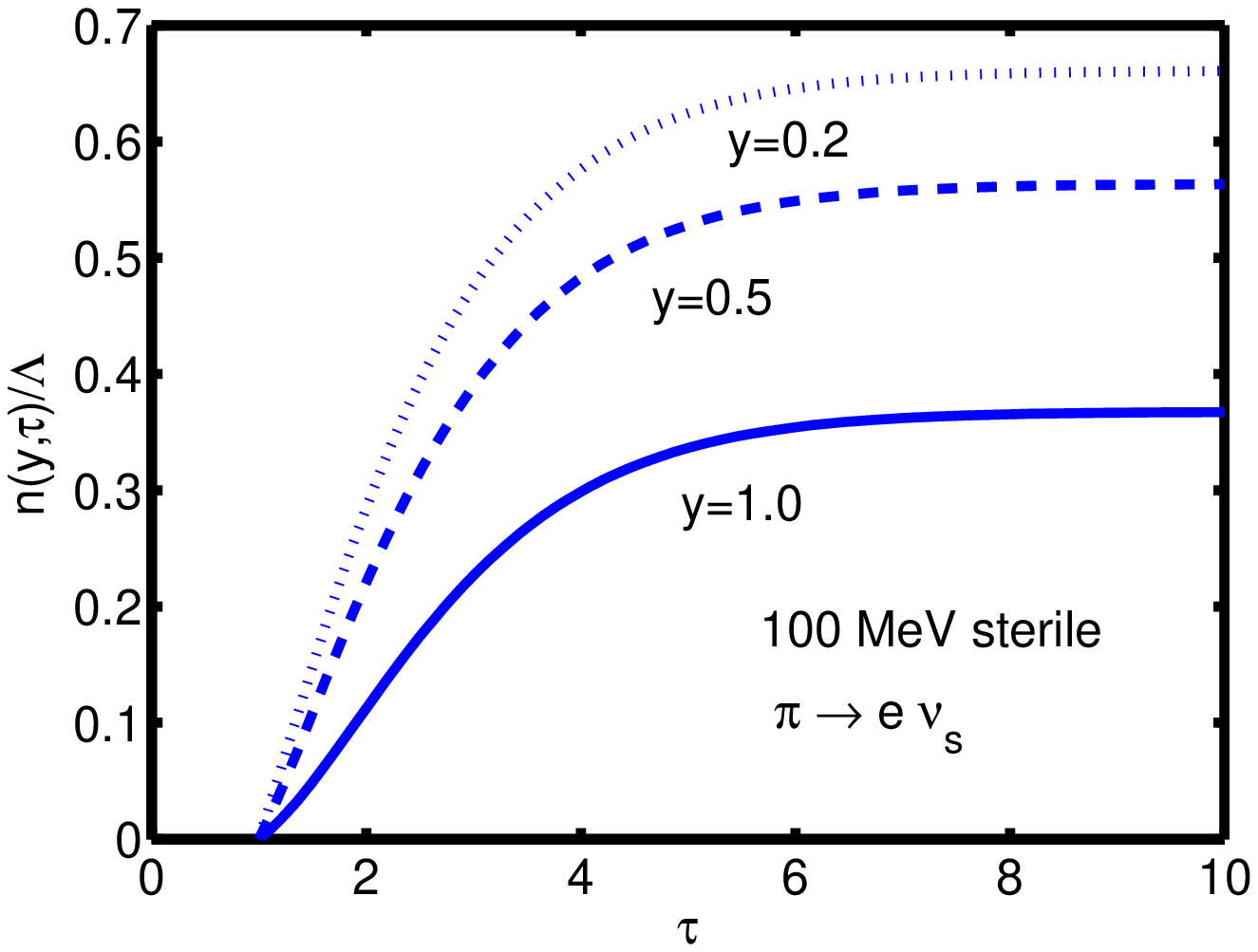}
\caption{Rates and population build up of a sterile $\nu_s$ obtained from quantum kinetics from $\pi \rightarrow l \nu_s$ with $l=\mu,e$ and $m_{\nu_s}$ near the kinematic threshold (for $\mu/e$ production $m_{\nu_s} = 30/100 MeV$ respectively).}
\label{fig:threshkinetics}
\end{center}
\end{figure}

 The rate equation and the resulting population buildup as a function of $\tau$ is shown in figures \ref{fig:evkinetics}-\ref{fig:threshkinetics} for several values of $y$ and $m_{\nu_s}$ for both $\pi \rightarrow \mu \nu_s$ and $\pi \rightarrow e \nu_s$. Note that the rate is enhanced for small values of y and is highly suppressed for large values of y. Fig \ref{fig:mevkinetics} clearly illustrates that freezeout occurs by $\tau=10$, which corresponds to temperatures $T \sim 15 MeV$, for a very wide range of sterile neutrino masses.

 A rough estimate of the sterile neutrino decoupling temperature can be made by considering the pion distribution. As the plasma temperature cools to well below the pion mass, the pion distribution will go as $f_{\pi} = e^{-m_{\pi}/T(t)}$ leading to a large suppression of the production rate at $T \lesssim 10 MeV$ which is when we expect the sterile neutrinos to freeze out. This is indeed what is found numerically in the population build up calculations of figs \ref{fig:evkinetics},\ref{fig:mevkinetics},\ref{fig:threshkinetics}.

\subsection{Light mass limit} \label{reldist}

As discussed, we expect freeze out to occur on the order of $T_f \sim 10-15 \mbox{MeV}$ and we will consider here light sterile neutrinos with $m_{\nu_s} \lesssim \mathcal{O}(MeV)$. Restricting attention to this mass range sets  $m_{\nu}/T_f \ll 1$  for the duration of sterile neutrino buildup and simplifies the kinetic equation considerably. For this particular production mechanism, it follows that $m_{\nu}^2 \ll m_{\pi}^2 - m_{l}^2$ and we introduce the parameter

\be
\Delta(m_{\nu}) \equiv \frac{m_{\pi}^2}{m_{\pi}^2-m_{l}^2 +m_{\nu}^2}
\ee so that, upon expanding in small parameters $m_{\nu}/T_f$ and $m_{\nu}^2/(m_{\pi}^2 -m_l^2)$ leads to the following simplifications

\be
\frac{E^{\pi}(p^+)}{T} = \frac{1}{\Delta}\frac{m_{\pi^2}}{m_{\nu}^2} y ~~;~~ \frac{E^{\pi}(p^-)}{T} = \Delta y + \frac{\tau^2}{4\Delta y}
\ee which is relevant for a wide range of light steriles that freeze out at $m_{\nu}/T_f \ll 1$. Note that we are suppressing the $m_{\nu}$ dependence of $\Delta$ and will do so for the remainder of this work (for $m_{\nu} \lesssim 1 MeV$). In this limit, the kinetic equation simplifies to

\be
\frac{1}{\Lambda}  \frac{dn}{d\tau} = \left(\frac{\tau}{y}\right)^2 \frac{ (1- \frac{m_{\pi}^2/6 f^2_\pi}{\tau^2})}{\left( e^{y}+1\right) } \ln \left( \left[ \frac{  1-e^{-\frac{m_{\pi}^2}{\Delta m_{\nu}^2}y} }{e^{-y} + e^{-\frac{m_{\pi}^2}{\Delta m_{\nu}^2}y}} \right]
\left[ \frac{ e^{-y} + e^{-\Delta y -\frac{\tau^2}{4\Delta y}  }}{1-e^{-\Delta y -\frac{\tau^2}{4\Delta y} } } \right]  \right) .
\ee We must ensure that the rate remain small in order to ignore the sterile's population build up and consequent Pauli blocking. The population scales with $\Lambda$ and if one were to compute the next order correction by including the first order buildup in the rate equation, the higher order correction would scale as $\Lambda^2$ and so on. Provided $\Lambda \ll 1$ (discussed below), the first order correction will be sufficient and higher order perturbations will be calculated in future work.

In order to evaluate the integral analytically, several mild simplifications are made. As previously mentioned, we use the fact that $g(t)$ varies slowly during the production process and a reasonable estimate is to instead use its average value (12.5). Additionally, if we are restricting our attention to the study of sterile neutrinos with $m_{\nu} \lesssim 1 MeV$, then the first bracketed term inside of the logarithm (which is independent of $\tau$) simplifies considerably.

\be
\frac{1}{\Lambda}  \frac{dn}{d\tau} = \left(\frac{\tau}{y}\right)^2 \frac{ (1- \frac{m_{\pi}^2/6 f^2_\pi}{\tau^2})}{\left( e^{y}+1\right) } \ln \left(
\frac{ 1 + e^{-(\Delta-1) y -\frac{\tau^2}{4\Delta y}  }}{1-e^{-\Delta y -\frac{\tau^2}{4\Delta y} } } \right)\,.
\ee The remaining $\tau$ dependence in the logarithm is a result of the Bose-Einstein suppression of the pions' thermal distribution and the $1/y^2$ dependence is a result of the phase space factors (with $m_{\nu} \lesssim 1 MeV$).

With these simplifications and by expanding the logarithms in a power series the integral can be carried out analytically. The final result is given as

\bea
 n(\tau,\tau_0,y) & = & \frac{\Lambda}{y^2(e^y+1)} \Bigg\{ \sum_{k=1}^{\infty} \Big[ (-1)^{k+1} e^{-(\Delta-1)ky} + e^{-\Delta k y} \Big] *  \\
   \Bigg[ \frac{4 \Delta^{3/2} y^{3/2} }{k^{5/2}}  \Bigg(\Gamma \left(\frac{k \tau_0^2}{4\Delta y}, \frac{3}{2}\right)
    & - & \Gamma\left(\frac{k \tau^2}{4\Delta y}, \frac{3}{2}\right) \Bigg)
     -  \, \frac{m_{\pi}^2 \Delta^{1/2} y^{1/2}}{6 f_{\pi}^2 k^{3/2}} \left( \Gamma\left(\frac{k \tau_0^2}{4\Delta y}, \frac{1}{2}\right) - \Gamma\left(\frac{k \tau^2}{4\Delta y}, \frac{1}{2}\right)\right) \Bigg] \Bigg\} \nonumber
\eea where $\Gamma(z,\nu)$ is the upper incomplete gamma function.

To get the frozen distribution, we take the long time limit, $\tau \rightarrow \infty$, to arrive at

\bea
 n(\tau,\tau_0,y) \Big|_{\tau \rightarrow \infty} & = & \frac{\Lambda}{y^2(e^y+1)} \Bigg\{ \sum_{k=1}^{\infty} \Big[ (-1)^{k+1} e^{-(\Delta-1)ky} + e^{-\Delta k y} \Big]   \\
   & * & \Bigg[ \frac{4 \Delta^{3/2} y^{3/2} }{k^{5/2}} \Gamma \left(\frac{k \tau_0^2}{4\Delta y}, \frac{3}{2}\right)
     -  \, \frac{m_{\pi}^2 \Delta^{1/2} y^{1/2}}{6 f_{\pi}^2 k^{3/2}} \Gamma\left(\frac{k \tau_0^2}{4\Delta y}, \frac{1}{2}\right) \Bigg] \Bigg\} \nonumber
\eea which can be written in a slightly different manner

\bea \label{frozen}
 n(\tau,\tau_0,y)\Big|_{\tau \rightarrow \infty} & = & f_d(\tau_0,y) = \frac{\Lambda}{y^2(e^y+1)} \sum_{k=1}^{\infty} \Big[ 1 + (-1)^{k+1}e^{ky} \Big] \frac{e^{-k \Delta y}}{k} J_k(\tau_0,y) \nonumber \\
  J_k(\tau_0,y) & = & 2 \tau_0 \left(\frac{\Delta y }{k}\right) e^{-k \tau_0^2/4 \Delta y} + \left( \frac{\Delta y}{k} \right)^{1/2}\Big[ \frac{2\Delta y}{k} - \frac{m_{\pi}^2}{6 f_{\pi}^2} \Big] \Gamma \left(\frac{k\tau_0^2}{4 \Delta y},\frac{1}{2}\right) \nonumber \,. \\
  & &
\eea

Eq \ref{frozen} is the decoupled distribution function of sterile neutrinos with $m_{\nu} \lesssim 1 MeV$ arising from pion decay near the QCD phase transition. This distribution function is valid for a wide range of sterile neutrino masses as we have only assumed $m_{\nu}/T(t) \ll 1$ for the period of production/freezeout ($ T_f \sim 10-15 MeV$), which is valid as long as we consider $m_{\nu} \lesssim 1 MeV$.

\begin{figure}[h!]
\includegraphics[height=4.0in,width=4.0in,keepaspectratio=true]{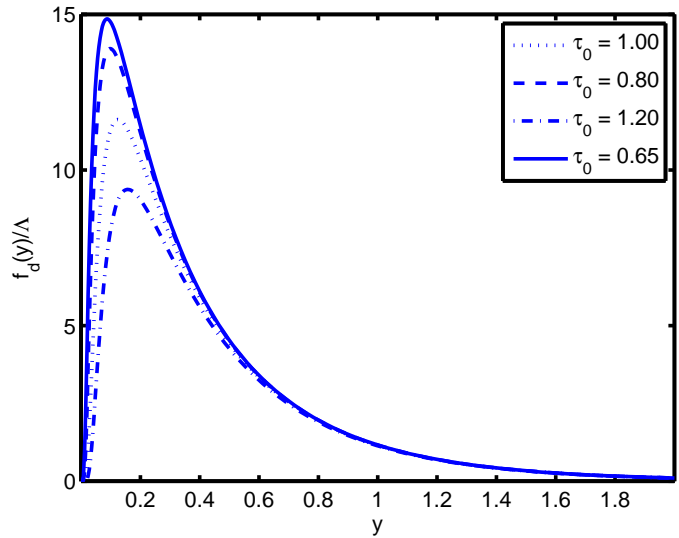}
\caption{The distribution function with various values of initial time. Note that an earlier initial time provides an enhancement with respect to later times.}
\label{fig:initialtimes}
\end{figure}

Note that the distribution function depends on the lower limit $\tau_0$. The distribution function is plotted for several values of $\tau_0$ in figure \ref{fig:initialtimes} where it can be seen that decreasing the lower limit increases the value of the distribution function. This is interpreted quite simply: production of steriles begins sooner and so the overall population is larger. If there are pions present in the plasma prior to the hadronization transition then this could be extended back to temperatures until the finite temperature corrections to the pion decay constant are no longer reliable: $\tau \sim m_{\pi}/\sqrt{6}f_{\pi} \sim 0.623$.

To see how this distribution differs from thermal distributions it is instructive to take the $y \rightarrow \infty$ and $y \rightarrow 0$ limits. Using that $\Gamma(k\tau_0^2/4\Delta y, 1/2) \rightarrow \Gamma(1/2) = \sqrt{\pi}$ as $y \rightarrow \infty$ gives the asymptotic form

\be
\frac{f_d(y,\tau_0)}{\Lambda} \Big|_{y \rightarrow \infty} = 2 \sqrt{\pi} \Delta^{3/2} \sum_{k=q}^{\infty} \left(\frac{1+ (-1)^{k+1}e^{ky}}{k^{5/2}}\right) \frac{e^{-(1+\Delta)ky}}{y^{1/2}} + \mathcal{O}\left(\frac{1}{y^{3/2}}\right) \rightarrow 0 \,.
\ee Taking the other limit $y \rightarrow 0$ along with use of the limiting expression of the $\Gamma$ function, $\Gamma(x,\nu)\big|_{x\rightarrow \infty} = x^{\nu}e^{-x} $, gives the asymptotic form for $y \rightarrow 0$

\be
\frac{f_d(y,\tau_0)}{\Lambda}\Bigg|_{y\rightarrow 0} = \sum_{k=1}^{\infty} \frac{ \left( 1+(-1)^{k+1}e^{ky} \right) }{k} \left( \frac{2 \Delta}{k} - \frac{m_{\pi}^2}{12 f_{\pi}^2 y} \right) \frac{\tau_0}{y} e^{-\frac{k\tau_0^2}{4 \Delta y}} \rightarrow 0 \,.
\ee Both of these asymptotic forms vanish but differ widely from the asymptotic forms of thermal distributions. This serves to illustrate the highly non-thermal nature of this distribution function.

The origin of the peak in this distribution becomes clearer with these insights. At low momentum, there is a competition between the phase space factor, $1/y^2$, and the thermal pion suppression, $e^{-\tau_0^2/4 \Delta y}$, which has a maximum at $y \sim \tau_0^2/4 \Delta$. A low momentum enhancement occurs in the Shi-Fuller mechanism as a result of a non-zero lepton asymmetry whereas the distribution considered here features similar low momentum enhancement from a combination of thermal suppression and phase space enhancement \emph{without the presence of a lepton asymmetry}.

Keeping the first term in the sum, $k=1$, provides an excellent approximation to the exact result with errors of only $~1\%$. With this approximation, the frozen distribution can be written as

\be
 f_d(\tau_0,y) = \frac{\Lambda}{y^2}  e^{-\Delta y} \left( 2 \tau_0 \left(\Delta y \right) e^{- \tau_0^2/4 \Delta y} + \left( \Delta y \right)^{1/2}\Big[ 2\Delta y - \frac{m_{\pi}^2}{6 f_{\pi}^2} \Big] \Gamma \left(\frac{\tau_0^2}{4 \Delta y},\frac{1}{2}\right) \right)
\ee where the second term is related to the error function via $\Gamma(x,1/2) = \sqrt{\pi}(1-\mbox{erf}(\sqrt{x}))$. Note that the approximate form features a maximum for $y \simeq 1/4\Delta$ (for $\tau_0 \simeq 1$) which is confirmed numerically. This approximation is discussed below.

\subsection{Ranges of validity}

\begin{figure}[h!]
\includegraphics[height=4.0in,width=4.0in,keepaspectratio=true]{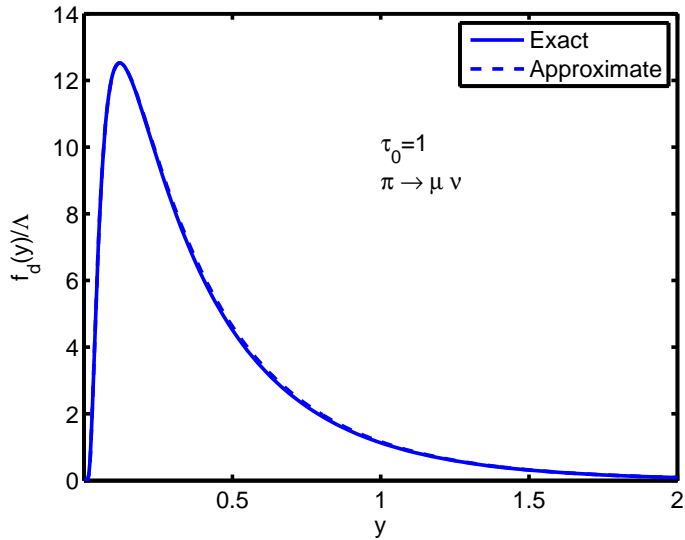}
\caption{The exact distribution function for small mass sterile neutrinos and an approximation keeping only the first term in the series expansion.}
\label{fig:approx}
\end{figure}

For light mass steriles, keeping just the first term in \ref{frozen} is an excellent approximation. In figure \ref{fig:approx} we have plotted both the exact distribution function and the first term of eq \ref{frozen}. Note that the two are nearly indistinguishable with errors only of about 1\%. This approximation can be understood simply because the higher order terms in the sums ($k > 1$) feature even more exponential suppression at both small and large momentum as seen in the asymptotic expressions.

The production process begins after $T_{QCD} \sim 155 MeV$ and is complete near $T \sim 10-15 MeV$ when the distribution freezes out. In terms of effective relativistic degrees of freedom, this implies starting with $g(T) \sim 14.25$ and concluding with $g(T) \sim 10.75$. As mentioned previously, g varies slowly which is seen in \cite{pdg} so the approximation replacing $g(T)$ with its average value $\bar{g} \sim 12.5$ is reasonable.

The approximation that the neutrino population can be neglected in the quantum kinetic equation requires that $\Lambda \ll 1$. This condition arises because upon iterating the first order solution (where the population was neglected) back into the kinetic equation would result in a perturbative expansion so that the rate equation would be of the form
\be
\frac{dn}{d\tau} \sim \sum_{n=1}^{\infty} \mathcal{O}(\Lambda)^{n} \,.
\ee If it were the case $\Lambda \sim \mathcal{O}(1)$ then our approximations break down and the kinetic equation would require the inclusion of higher order processes. Using the values from ref \cite{pdg}, the dimensionless scale $\Lambda$ can be written as
\be
\Lambda(T) = 6.511 \left(\frac{m_{\pi}^2}{GeV^2}\right) \left( \frac{|U_{ls}|^2}{10^{-5}}\right) \left( \sqrt{\frac{12.5}{g(T)} }\right) \left( \frac{m_l^2 +m_{\nu_s}^2}{m_{\pi}^2} -\frac{(m_l^2-m_{\nu_s}^2)^2}{m_{\pi}^4} \right)
\ee which clearly depends on the value of the sterile neutrino mass.

Taking $m_{\nu} \lesssim 1 MeV$ implies that for $\pi \rightarrow \mu \nu$ and $\pi \rightarrow e \nu$, the neutrino mass may be neglected in the expressions for $\Lambda$. For this situation, the parameters reduce simply to

\be
\Lambda_{\mu} \lesssim 0.03 |U_{\mu s}|^2/10^{-5} ~~;~~ \Lambda_e \lesssim 0.5|U_{es}|^2
\ee so that, to leading order, neglecting the sterile population is a good approximation for small mixing.

We had investigated light sterile neutrinos with $m_{\nu_s} \le 1 MeV$ and in this range the distribution function varies negligibly with $m_{\nu}$. If we want to consider sterile neutrinos with $m_{\nu} \gtrsim 1 MeV$, the approximations made previously will break down and a full numerical evaluation of the rate equation will be needed. The focus on heavy sterile neutrinos and the effect on cosmological measurements will be the study of forthcoming work where we expect nontrivial deviations from the results presented here.

\section{Observational consequences}

\subsection{Bounds from dark matter and dwarf spheroidals}

The sterile neutrino energy density today is given by Eq \ref{energydensity}. Note that freezeout occurs between $\tau \sim 3-5$ or $T \sim 10-15 MeV$, so that, as mentioned in the previous section, the particles are relativistic at the time of decoupling. For the light sterile neutrinos we consider here, we relate the number of relativistic species at the time of sterile decoupling to the photon temperature today by the usual relation between the plasma and photon temperatures:
\be
T_{plasma}(z=0) = \left(\frac{2}{g_d}\right)^{1/3}T_{\gamma,0} \lesssim 10^{-4}eV  ~~;~~ T_{\gamma,0} = 2.35*10^{-4} eV\,.
\ee For $m_{\nu} \gtrsim 0.01 eV$, we may neglect the $(y/x)^2$ term in eq \ref{energydensity} ($\rho$) and therefore the sterile neutrinos are non-relativistic \emph{today}:

\be
\rho_{\nu,0} = g_{\nu} m_{\nu} \frac{2}{g_d} \frac{T_{\gamma,0}^3}{2 \pi^2}  \int dy \, y^2  f_d(q_c) = m_{\nu} n_{\nu}(t_0)  \,.
\ee With this, the contribution to the density today is obtained using the distribution function calculated in section \ref{reldist} and eq \ref{omegadm} to give

\be
\frac{\Omega_{\nu_s,0}}{\Lambda} = \frac{h^2 n_{\gamma}}{\rho_c} \frac{g_{\nu} m_{\nu}}{2 \zeta (3) g_d} \int_0^{\infty} dy \, y^2 \frac{f_d(y)}{\Lambda} \equiv \frac{h^2 n_{\gamma}}{\rho_c} \frac{g_{\nu} m_{\nu}}{2 \zeta (3) g_d} I_0 (m_{\nu})
\ee where

\be
I_{n}( m_{\nu}) = \int_0^{\infty} dy \, y^{2+n} \frac{f_d(y)}{\Lambda} \,.
\ee When $m_{\nu} \lesssim 1 MeV$ the moments, $I_{n}(m_{\nu})$ do not vary significantly and, for this mass range, they may be approximated by their value at $m_{\nu} = 0$. We work under the assumption that $m_{\nu} \lesssim 1 MeV$ which so that we may use the limit $I_{n}(0)$ in subsequent calculations and the limiting values are listed in table \ref{tab:limittable}.

\begin{table}[h]
\caption{Table of limiting values for the function $I_{n}(0)$.} \label{tab:limittable}
\begin{tabular}{|c|c|c|c|}
\hline
\multicolumn{4}{|c|}{ $I_{n}(0)~~;~~\pi\rightarrow l \nu$ }   \\ \hline
\backslashbox{$ l $}{$ n $}  & 0 & 1 & 2 \\ \hline
 $ e$ & 3.756 & 9.675 &  34.300  \\ \hline
 $\mu$ &  1.830  & 2.140 & 3.426  \\ \hline
\end{tabular}
\end{table}

Using the results of sec \ref{decoupleddynamics}, if we consider sterile masses with $m_{\nu_s} \ll m_l$ then we may neglect the sterile mass in both $\Delta, \Lambda$ so that we arrive at

\be
\frac{\Omega_{\nu_s,0} h^2}{\Lambda} = \frac{h^2 n_{\gamma}}{\rho_c} \frac{g_{\nu} m_{\nu}}{2 \zeta (3) g_d} I_0 (0)\ee Considering light scalars simplifies the scales, $\Lambda$, so that the appropriate scales in the problem are
\be
\Lambda_{\pi \rightarrow l \nu}(m_{\nu} = 0) \equiv  \Lambda_l ~~;~~ \Lambda_{\mu} = 0.032*\frac{|U_{\mu s}|^2}{10^{-5}} ~~;~~ \Lambda_e = 1.7*10^{-6}*\frac{|U_{es}|^2}{10^{-5}}
\ee so that

\be
m_{\nu_s} \Lambda \le  \frac{ \Omega_{DM} h^2}{n_\gamma h^2 / \rho_c} \left(\frac{g_d}{g_{\nu_s}}\right) \frac{2 \zeta(3)}{I_0(0)}
\ee Using the values from \cite{pdg} of $ n_{\gamma} h^2 /\rho_c = 1/25.67eV$ and $\Omega_{DM} h^2 = 0.1199$ while assuming $g_{\nu_s} = 2$ and $g_d = \bar{g} = 12.5$ leads to the bounds

\be
m_{\nu_s} \frac{|U_{\mu s}|^2}{10^{-5}} \le 0.739 keV ~~;~~ m_{\nu_s} \frac{|U_{e s}|^2}{10^{-5}} \le 7242 keV \,. \label{highlim}
\ee

As discussed in sec \ref{decoupleddynamics}, the dark matter phase space density decreases over the course of galactic evolution and the primordial phase space density may be used as an upper bound to obtain limits on the mass of dark matter. Using observational values for dwarf spheroidal galaxies from ref \cite{destri} a set of bounds complementary to those from CMB measurements can be obtained. As discussed, imposing the condition $\mathcal{D}_{p} \ge \mathcal{D}_{nr}$ gives us the constraint

\be
\mathcal{D}_{p} \ge \frac{1}{3^{3/2} m_{\nu_s}^4} \left. \frac{\rho}{\sigma^3} \right|_{today}
\ee Assuming, as before, that the sterile neutrino mass is much smaller than the charged lepton mass renders the phase space density independent of the sterile neutrino mass. This leads to a bound on the mass given as

\be
m_{\nu_s} \ge \left[\frac{1}{3^{3/2}} \frac{\rho}{\sigma^3} \Big|_{today} \mathcal{D}_p^{-1} \right]^{1/4} \,.
\ee Using the results from section \ref{decoupleddynamics}, the phase space density is given as

\be
\mathcal{D} = \frac{g_{\nu_s} \Lambda}{2\pi^2}\frac{I_0(0)^{5/2}}{I_{2}(0)^{3/2}}
\ee so that the bound becomes

\be
m_{\nu_s} \Lambda^{1/4} \ge \left( \frac{2 \pi^2}{3^{3/2} g_{\nu_s}} \frac{\rho}{\sigma^3} \Big|_{today} \frac{I_2(0)^{3/2}}{I_0(0)^{5/2}} \right)^{1/4}
\ee which can serve as a complementary bound to the limits set from $\Omega_{DM}$. Values of the phase space density today are summarized in ref \cite{destri} and using the data from the most compact dark matter haloes leads to bounds on sterile neutrino dark matter. The halo radius ($r_h$), velocity dispersion ($\sigma$), phase space density today and the calculated bounds are summarized in table \ref{tab:phasespace} where we chose several of the most compact dwarf spheroidal galaxies (a more thorough list is available in \cite{destri}).

\begin{figure}[h!]
\begin{center}
\includegraphics[height=3.5in,width=3.2in,keepaspectratio=true]{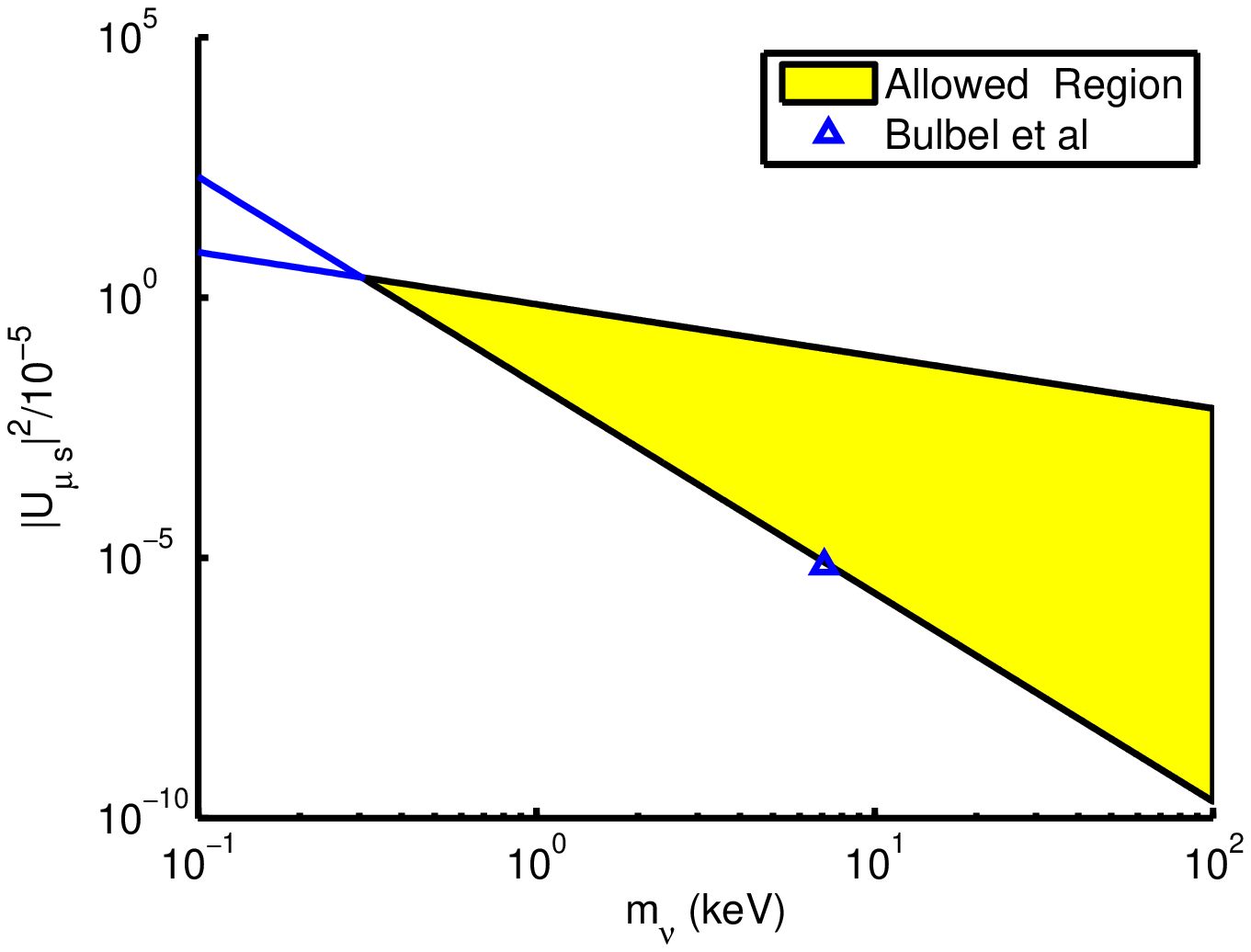}
\includegraphics[height=3.5in,width=3.2in,keepaspectratio=true]{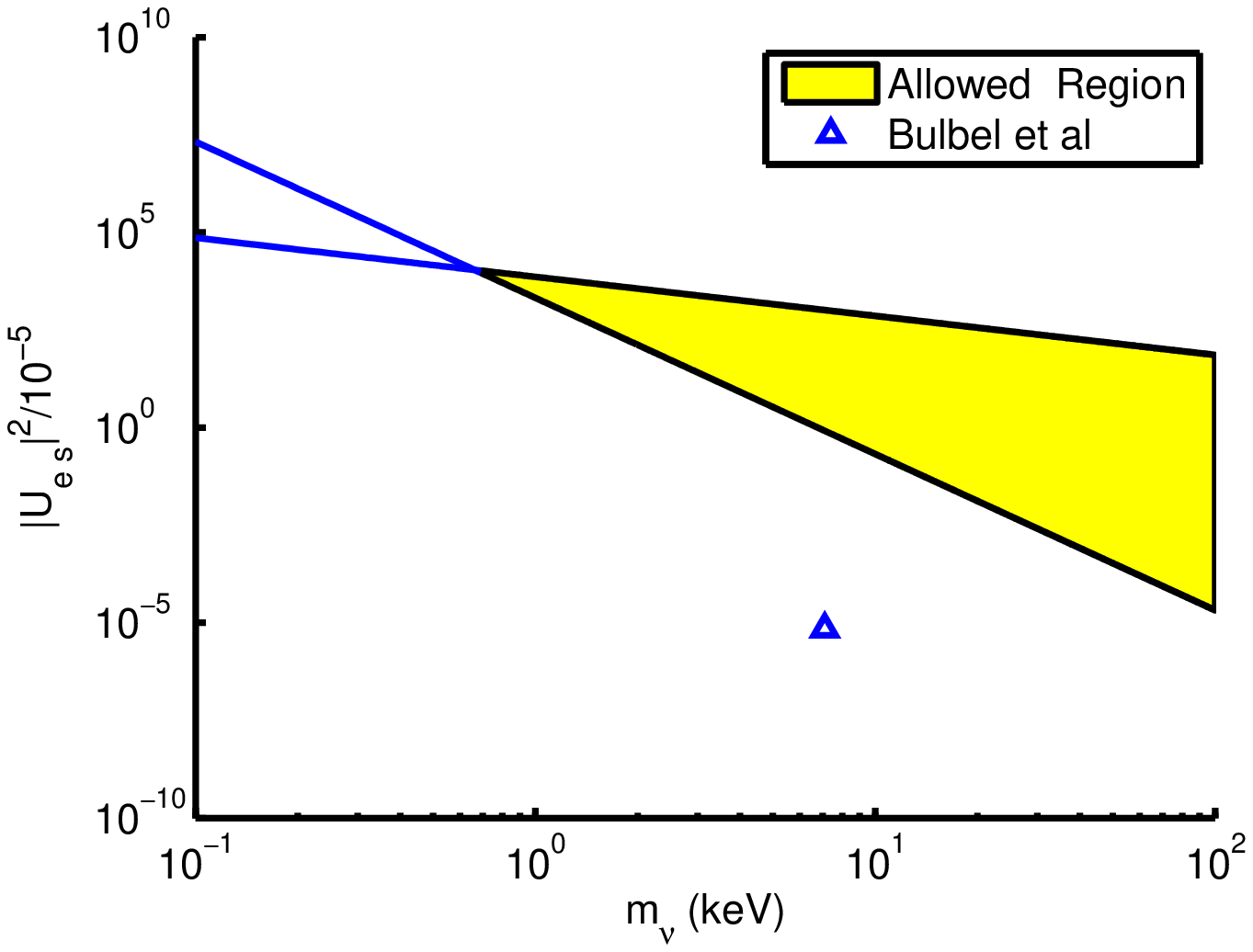}
\caption{The bounds on sterile mass and mixing obtained from CMB and galactic measurements. The allowed regions determined from Eqs \ref{highlim},\ref{lowlim} are shaded and the sterile neutrino parameters which potentially explain the 3.5 keV signal (Bulbul et al) are also shown.}
\label{fig:bound}
\end{center}
\end{figure}

\begin{table}[h]
\caption{Phase space data for compact galaxies and derived bounds on sterile neutrinos arising from pion decay.} \label{tab:phasespace}
\begin{tabular}{|c|c|c|c|c|c|}
\hline
Galaxy  & $\frac{r_h}{pc}$ & $\frac{\sigma}{km/s}$ & $\frac{\rho/\sigma^3}{(keV)^4}$ & $\frac{ m_{\nu} \Lambda^{1/4}_{\mu}}{keV} \big|_{\min}$ & $\frac{ m_{\nu} \Lambda^{1/4}_{e}}{keV} \big|_{\min}$ \\ \hline
 Willman 1 & 19  & 4 & 0.723 & 1.178 & 1.782 \\ \hline
 Segue 1   & 48  & 4 & 1.69 & 1.456  & 2.204    \\ \hline
 Coma-Berenices & 123  & 4.6 & 0.04 & 0.571 & 0.864  \\ \hline
 Leo T & 170  & 7.8 & 0.014 & 0.4392 & 0.665 \\ \hline
 Canis Venatici II & 245  & 4.6 & 0.04 & 0.571  & 0.864 \\ \hline
 Draco & 305  & 10.1 & 0.0036  & 0.3128 & 0.473 \\ \hline
 Fornax & 1730 & 10.7 & 2.56*$10^{-4}$ & 0.1615 & 0.2445 \\ \hline
\end{tabular}
\end{table} Taking the minimum value from this data set translates into the bounds

\be
m_{\nu} \left(\frac{|U_{\mu s}|^2}{10^{-5}}\right)^{1/4} \ge 0.38 \mbox{keV} ~~;~~ m_{\nu} \left(\frac{|U_{e s}|^2}{10^{-5}}\right)^{1/4} \ge 6.77 \mbox{keV} \,. \label{lowlim}
\ee The bounds from dwarf galaxies can be combined with the bounds from CMB measurements of $\Omega_{DM}$ to obtain allowed regions of parameter space. The two bounds are illustrated in Fig. \ref{fig:bound} along with the parameter values reported in ref. \cite{bulbul} arising from the 3.5 keV x-ray signal. If sterile neutrinos are responsible for the x-ray signal then production from $\pi \rightarrow \mu \nu$ is a mechanism consistent with the data within a narrow region while sterile neutrinos produced from $\pi \rightarrow e \nu$ are not.

\subsection{Equation of State and Free streaming}

The equation of state for an arbitrary dark matter candidate is characterized by the parameter $w(T)$ given by eq \ref{eos}. A light sterile neutrino ($m_{\nu} \lesssim 1 MeV$) freezes out while it is still relativistic since $m/T \ll 1$ during production/freezeout therefore the results of the previous section hold. This distribution will then determine at what temperature this species becomes non relativistic via Eq \ref{eos}, which is rewritten here explicitly in terms of $m_{\nu}/T$:

\be
w(T) = \frac{\mathcal{P}}{\rho} = \frac{1}{3} \frac{ \int dy \, \frac{y^4}{ \sqrt{y^2+\frac{m_{\nu}^2}{T(t)^2}}} f_d(q_c)}{\int dy \, y^2 \sqrt{y^2+\frac{m_{\nu}^2}{T(t)^2}} f_d(q_c)} \,.
\ee

\begin{figure}[h!]
\begin{center}
\includegraphics[height=3.2in,width=3.0in,keepaspectratio=true]{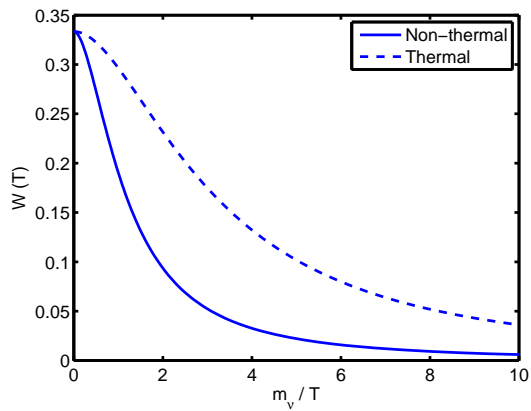}
\caption{Equation of state compared to thermal. }
\label{fig:eos}
\end{center}
\end{figure}

Many fermionic dark matter candidates which freeze out at temperature $T_f$ are treated as being in LTE in the early universe so that their distribution functions are given by the standard form

\be
f_{LTE}(y) = \frac{1}{e^{\sqrt{y^2+ m^2/T_f^2}} + 1} \,.
\ee To compare the new distribution to thermal results, assume that thermal particles with the same mass also freezeout while relativistic. The equation of state arising from thermal distributions and the non-thermal distribution we obtain are plotted as a function of $m_{\nu}/T$ in fig \ref{fig:eos}. Note that the non-thermal distribution equation of state parameter is smaller for all times. This is a reflection of the enhancement of small momentum so that the non-thermal distribution results in a dark matter species which is colder and becomes non relativistic much earlier than the thermal result. In summary, the thermal distribution produces particles that become non-relativistic when $m/T \gg 1$ whereas the pion decay mechanism produces particles that become non-relativistic when $m/T \sim 1$. This non-thermal distribution function produces a dark matter candidate that is \emph{colder} than those produced at LTE.

The free streaming wave vector enters when one considers a linearized collisionless Bolzmann-Vlasov equation describing the evolution of gravitational perturbations which ultimately lead towards structure formation \cite{boyanfreestream,moredevega}.  The free streaming wave vector $k_{fs}$ leads to a cutoff in the linear power spectrum of density perturbations and is given by
\be
k^2_{fs} =  \frac{ 4 \pi G \rho}{ \overline{ \vec{V}^2}} \,.
\ee Modes with $k<k_{fs}$ lead to gravitation collapse in a manner akin to the Jeans instability. This is shown explicitly and discussed at length in ref \cite{boyanfreestream}. Assuming that a light sterile neutrino is the only dark matter (so that $\rho_{\nu_s} = \rho_{DM}$) and using the results of section \ref{decoupleddynamics} (for a non-relativistic species), the free streaming wave vector is given by

\be
k^2_{fs}  = \frac{3}{2} \frac{\Omega_{DM} H^2}{\overline{ \vec{V}^2}} = \frac{3}{2} H^2 \Omega_{DM} \left( \frac{m_{\nu}}{T(t)} \right)^2 \frac{\int dy \, y^2 f_d(y)}{\int dy \, y^4 f_d(y)} \,.
\ee Using the latest values from Planck \cite{planck} sets the free streaming wave vector as

\be
k_{fs}(z=0) = \frac{m_{\nu}}{T_{\gamma,0}} \left( \frac{g_d}{2} \right)^{1/3} \sqrt{\frac{3}{2} \Omega_{DM,0} H_0^2 \frac{I_0(0)}{I_2(0)}} = \frac{0.617}{kpc} \frac{m_{\nu}}{keV} \left(\frac{g_d}{2}\right)^{1/3} \sqrt{\frac{I_0(0)}{I_2(0)}}
\ee or in terms of the free streaming length, $\lambda_{fs} = 2\pi/k_{fs}$,

\be
\lambda_{fs}(0) = 10.2 kpc \, \left(\frac{keV}{m_{\nu}}\right) \left(\frac{2}{g_d}\right)^{1/3} \sqrt{\frac{I_2(0)}{I_0(0)}}
\ee For a redshift z during matter domination the free streaming length scales as $\lambda_{fs}(z) = \lambda_{fs}(0)/\sqrt{1+z}$ and the free streaming length today for the particular processes are then given by

\be
\lambda_{fs}^{\mu} (0) = 7.6 \mbox{kpc} \left(\frac{\mbox{keV}}{m_{\nu}}\right)         ~~;~~ \lambda_{fs}^e(0) =16.7 \mbox{kpc} \left(\frac{\mbox{keV}}{m_{\nu}}\right)
\ee where we've used the notation $\lambda_{fs}^l(0) \equiv \lambda_{fs}(0) \big|_{\pi \rightarrow l \nu}$.

\subsection{Contributions to Dark Radiation}

In previous sections we considered sterile neutrinos with $m_{\nu} \lesssim 1 MeV$ specifically with $m_{\nu} \sim keV$ in mind.  As discussed in sec \ref{decoupleddynamics}, cosmological measurements can directly probe additional neutrino species through the number of effective relativistic species. We have argued that the sterile neutrinos under consideration in this work will decouple while relativistic at temperatures on the order of $10-15 MeV$ and will remain relativistic until $T \sim m_{\nu}$.

In order to contribute to $N_{eff}$, a sterile neutrino must have mass $m_{\nu} \lesssim 1 eV$ so that it remains relativistic through matter-radiation equality. The previous general analysis still holds but here we consider specifically sterile neutrinos with $m_{\nu} \lesssim 1 eV$, those which are currently of interest for accelerator searches \cite{lsnd,miniboone}. The modifications to $N_{eff}$ with the sterile neutrinos produced from pion decay are given by Eq \ref{neff} and rewritten here as

\be
\Delta N_{eff} = \Lambda \frac{60 g_{\nu_s}}{7\pi^4} \left(\frac{11}{2 g_d}\right)^{4/3}I_{1}(m_{\nu}) \,.
\ee As mentioned, in order to contribute to $N_{eff}$, the neutrinos must remain be relativistic at the time of matter-radiation equality, $T \sim eV$, so this is only valid for $m_{\nu} \lesssim 1 eV$. In this range of masses, $I_1(m_{\nu})$ does not vary appreciably and is very nearly its value for $m_{\nu} = 0$ which is listed in table \ref{tab:limittable}. For the different processes we have

\be
 \Delta N_{eff}\Big|_{\pi \rightarrow \mu \nu} = 0.0040 *\frac{|U_{\mu s}|^2}{10^{-5}} ~~;~~ \Delta N_{eff} \Big|_{\pi \rightarrow e \nu} = 9.7 *10^{-7} \frac{|U_{e s}|^2}{10^{-5}} \,.
\ee The measurement from Planck is consistent with $\Delta N_{eff} \lesssim 0.4$ \cite{planck} and using bounds from land based experiments summarized in \cite{giunti2,mirizzi} we can get an estimate of whether these light sterile will contribute significantly.

Kamland and Daya Bay \cite{superk,dayabay} recently reported upper bounds of $|U_{\mu s}|^2 < 0.01$ for the mass squared difference $10^{-3} eV^2 < |\Delta m_{1 s}|^2 <0.1 eV^2$. Taking the upper bound leads to $\Delta N_{eff} < 4$ suggesting that $\pi \rightarrow \mu \nu_s$ can contribute significantly to $N_{eff}$ for a $\sim 1 eV$ sterile. Ground based experiments which suggest $m_{\nu_s} \sim 1 eV$ could be in tension with CMB measurements which suggest $\Delta N_{eff} \lesssim 0.4$ and $m_{\nu_s} \lesssim 0.30 eV$ if the upper bound on the mixing is near its true value. Conversely, if $N_{eff}$ could be measured more accurately, this could potentially be used to place tighter bounds on $|U_{l s}|^2$. For instance, the latest results from the Planck collaboration suggest that $\Delta N_{eff} < 0.15$ \cite{planck2} which leads to the constraint $|U_{\mu s}|^2 < 3.8*10^{-4}$.

\section{Summary, Discussion and Further Questions }

We studied the production of sterile neutrinos from $\pi \rightarrow l \nu_s$ shortly after the QCD phase transition (crossover) in the early universe. Pions, being the lightest pseudoscalar mesons, are copiously produced through hadronization after the confinement-deconfinement and chiral phase transition at $T\simeq 155 MeV$ with their primary decay channel purely leptonic. Pions will be present in the plasma with a thermal distribution, maintaining LTE via strong, electromagnetic and weak interactions maintaining detailed balance (with charged leptons and active neutrinos) for kinetic and chemical equilibrium. However, pions will decay into sterile neutrinos via their mixing with active ones. We include finite temperature corrections to the pion mass and decay constant to assess the production properties of a sterile species via $\pi$ decay but in absence of a lepton asymmetry.

For sterile neutrino masses $\lesssim 1 MeV$ we find that they are produced with a highly non-thermal distribution function and freeze out at $T_f \simeq 10-15 MeV$. The distribution function features a sharp enhancement at low momentum resulting from a competition between phase space and thermal suppression of the parent meson. The strong low momentum enhancement featured in this non-thermal distribution function makes the species very \emph{cold} despite such a small mass, and is remarkably similar to that found in resonant production via a lepton asymmetry \cite{shifuller,abazajian2}; however, we emphasize that our study considered vanishing lepton asymmetry.

The frozen distribution function depends on a particular combination of the mass of the sterile neutrino and mixing matrix element $U_{ls}$. Dark matter abundance constraints from the CMB and constraints from the most dark matter dominated dwarf spheroidal galaxies provide upper and lower bounds respectively on combinations of $m_s, U_{ls}$. These bounds feature a region of compatibility with the recent observations of a $3.55 keV$ line that could imply a $7 keV$ sterile neutrino as dark matter candidate.

\bea
m_{\nu_s} \frac{|U_{\mu s}|^2}{10^{-5}} \le 0.739 \, \mbox{keV} & ; & m_{\nu_s} \frac{|U_{e s}|^2}{10^{-5}} \le 7242 \, \mbox{keV} \nonumber \\
m_{\nu} \left(\frac{|U_{\mu s}|^2}{10^{-5}}\right)^{1/4} \ge 0.38 \, \mbox{keV} & ; & m_{\nu} \left(\frac{|U_{e s}|^2}{10^{-5}}\right)^{1/4} \ge 6.77 \, \mbox{keV}
\eea

An important characteristic for structure formation is the free streaming wavevector and length, $k_{fs} = 2\pi/\lambda_{fs}$, where $k_{fs}$ determines a cutoff in the linear power spectrum of density perturbations and consequently $\lambda_{fs}$ determines the length scale below which gravitation collapse is suppressed. This scale is determined by the distribution function at freeze-out and the mass of the (non-relativistic) DM component. We find that the highly non-thermal distribution function from $\pi$ decay determines that this DM species is \emph{colder} with a $\lambda_{fs} \simeq \mbox{few kpc}$ today, consistent with the scale of cores observed in dwarf spheroidal galaxies. We find (today)

\be
\lambda_{fs}^{\mu} (0)^2 = 7.6 \, \mbox{kpc} \left(\frac{\mbox{keV}}{m_{\nu}}\right)         ~~;~~ \lambda_{fs}^e(0)^2 =16.7 \, \mbox{kpc} \left(\frac{\mbox{keV}}{m_{\nu}}\right)
\ee

If the mass of sterile neutrinos is $m_{\nu_s} < 1 eV$ they may contribute to the radiation component between matter radiation equality and photon decoupling thereby contributing to the effective number of relativistic degrees of freedom $N_{eff}$. The most recent accelerator and astrophysical bounds on the masses and mixing angles of sterile neutrinos in $3+1$ or $3+2$ schemes \cite{superk,dayabay,giunti2,mirizzi} combined with the result for the frozen distribution function suggest substantial contributions from this species to $N_{eff}$ although severe tensions remain between accelerator data and Planck bounds from the CMB

\be
 \Delta N_{eff}\Big|_{\pi \rightarrow \mu \nu} = 0.0040 *\frac{|U_{\mu s}|^2}{10^{-5}} ~~;~~ \Delta N_{eff} \Big|_{\pi \rightarrow e \nu} = 9.7 *10^{-7} \frac{|U_{e s}|^2}{10^{-5}} \,.
\ee \\

\textbf{Further Questions}

While we focused on ``light'' sterile neutrinos with $m_{\nu_s} < 1 MeV$, there are potentially important aspects to be studied for the case of $10 MeV \lesssim m_{\nu_s} \lesssim 140 MeV$, a range of masses kinematically available in $\pi \rightarrow e \nu_s$. These ``heavier'' species may actually contribute as a CDM component since freeze-out still occurs at a scale $T_f \sim 10-15 MeV$ therefore this species will be non-relativistic and \emph{non-thermal} upon freeze out. Heavy sterile neutrinos may decay into lighter active neutrinos on time scales larger than that for BBN. These late-produced active neutrinos would be injected into the cosmic neutrino background \emph{after} neutrino decoupling and will therefore not be able to reach LTE with the plasma becoming a non-LTE active neutrino component which may contribute to $N_{eff}$ non-thermally. We expect to report on these issues in further studies.

\acknowledgments D.B. and L.L. acknowledge support from NSF through grant PHY-1202227.

\end{document}